\documentclass[twocolumn,numbook,nospthms,natbib]{svjour3_arxiv}          
 
\smartqed  
 
\usepackage{graphicx}
\usepackage{amsmath,amssymb,amsthm} 
\usepackage{enumerate}
\usepackage{enumitem}
\usepackage{epstopdf}
\usepackage{color}
\usepackage{pdfsync}
\usepackage{float}
\usepackage{graphicx}
\usepackage{xspace}
\usepackage{algpseudocode}
\usepackage{algorithmicx}
\usepackage{algorithm}
\usepackage{subfigure}
\usepackage{multirow}

\usepackage{tikz}
\usetikzlibrary{positioning,arrows}
\tikzset{
block/.style={
  draw,
  rectangle,
  minimum height=0.8cm,
  minimum width=0.8cm, align=center
  },
line/.style={->,>=latex'}
}
\usetikzlibrary{arrows}

\newcommand{\cbrown}{\textcolor{black}}

\newcommand{\beq}{\begin{equation}}
\newcommand{\eeq}{\end{equation}}
\newcommand{\beqa}{\begin{eqnarray}}
\newcommand{\eeqa}{\end{eqnarray}}
\newcommand{\nn}{\nonumber}

\newcommand{\bxi}{{\mbox{\boldmath $\bxi$}}}

\newcommand{\rw}{\rightarrow}
\newcommand{\Prob}{\mathbb{P}}
\newcommand{\mbbQ}{\mathbb{Q}}
\newcommand{\mbbE}{\mathbb{E}}
\newcommand{\Real}{\mathbb{R}}

\newcommand{\mB}{{\mathcal B}}

\newcommand{\mF}{{\mathcal F}}

\newcommand{\mX}{{\mathcal X}}
\newcommand{\mY}{{\mathcal Y}}

\newcommand{\mP}{{\mathcal P}}

\newcommand{\x}{\textit{\textbf{x}}}
\newcommand{\y}{\textit{\textbf{y}}}

\newcommand{\s}{\textit{\textbf{s}}}
\newcommand{\X}{\textit{\textbf{X}}}
\newcommand{\Y}{\textit{\textbf{Y}}}

\newcommand{\mfL}{($\mathfrak{L}$)}
\newcommand{\mfD}{($\mathfrak{D}$)}
\newcommand{\mfC}{($\mathfrak{C}$)}

\newtheorem{Teorema}{\em Theorem}
\newtheorem{Definition}{\em Definition}
\newtheorem{Lema}{\em Lemma}
\newtheorem{Proposicion}{\em Proposition}

\newtheorem{Assumption}{\em Assumption}
\newtheorem{Nota}{\em Remark}

\begin{document}

\title{{On the performance of particle filters with adaptive number of particles}\thanks{This work was partially supported by \emph{Agence Nationale de la Recherche} of France under PISCES project (ANR-17-CE40-0031-01), the Office of Naval Research (award no. N00014-19-1-2226), {\em Agencia Estatal de Investigaci\'on} of Spain (RTI2018-099655-B-I00 CLARA), and NSF through the award CCF-1618999.}
}



\author{V\'ictor Elvira \and Joaqu\'in Miguez \and Petar M. Djuri\'c}


\institute{V. Elvira \at
	School of Mathematics,\\
	University of Edinburgh (UK).\\
	\email{victor.elvira@ed.ac.uk}
\and
	J.~Miguez \at
	Department of Signal Theory \& Communications,\\
	Universidad Carlos III de Madrid (Spain).\\
	IIS Gregorio Mara\~n\'on (Spain).\\
	\email{joaquin.miguez@uc3m.es} 
\and
	P. M. Djuri\'c \at
	Department of Electrical and Computer Engineering,\\ 
	Stony Brook University (USA).\\
	\email{petar.djuric@stonybrook.edu}
}

\date{\today} 

\maketitle

\begin{abstract}
\cbrown{ 
We investigate the performance of a class of particle filters (PFs) that can automatically tune their computational complexity by evaluating online certain predictive statistics which are invariant for a broad class of state-space models. To be specific, we propose a family of block-adaptive PFs based on the methodology of \citet{elvira2017adapting}. In this class of algorithms, the number of Monte Carlo samples (known as \textit{particles}) is adjusted periodically, and we prove that the theoretical error bounds of the PF actually adapt to the updates in the number of particles. The evaluation of the predictive statistics that lies at the core of the methodology is done by generating \textit{fictitious observations}, i.e., particles in the observation space. We study, both analytically and numerically, the impact of the number $K$ of these particles on the performance of the algorithm. In particular, we prove that if the predictive statistics with $K$ fictitious observations converged exactly, then the particle approximation of the filtering distribution would match the first $K$ elements in a series of moments of the true filter. This result can be understood as a converse to some convergence theorems for PFs. From this analysis, we deduce an alternative predictive statistic that can be computed (for some models) without sampling any fictitious observations at all. Finally, we conduct an extensive simulation study that illustrates the theoretical results and provides further insights into the complexity, performance and behavior of the new class of algorithms. 
}

\end{abstract}

\tolerance = 1600
\pretolerance = 1600

\section{Introduction}
\label{sec_introduction}
In science and engineering, there are many problems that are studied by way of dynamic probabilistic models. Some of these models  describe mathematically the evolution of hidden states and their relations with observations, which are sequentially acquired. In many of these problems, the objective is to estimate sequentially the posterior probability distribution of the hidden model states. A methodology that has gained considerable popularity in the last two and a half decades is particle filtering (also known as {sequential} Monte Carlo) {\citep{Gordon93,Liu98,Doucet01b,Djuric03,Kunsch13}}. This is a Monte Carlo methodology that approximates the distributions of interest by means of random (weighted) samples.

Arguably, a key parameter of particle filters (PFs) is the number of generated Monte Carlos samples (usually termed \textit{particles}). A larger number of particles improves the approximation of the filter but also increases the computational complexity. However, it is impossible to know a priori the appropriate number of particles to achieve a prescribed accuracy in the estimated parameters and distributions. \cbrown{So a question of great practical interest is how to determine the necessary number of particles to achieve a prescribed performance and, in particular, how to determine it automatically and in real time.}

\subsection{\cbrown{Particle filtering with time-varying number of particles}}
Until the publication of \citet{elvira2017adapting}, \cbrown{not many papers had considered the selection/adaptation of the number of particles in a systematic and rigorous manner}. In \citet{elvira2017adapting}, a methodology was introduced to address this problem with the goal of adapting the number of particles in real time. \cbrown{The method is based on a rigorous mathematical analysis and we discuss it in more detail in Section \ref{ssIntroBackground}.}  

Other efforts toward the same goal  include the use of a Kullback-Leibler divergence-based  approximation error by \citet{fox2003adapting}, where the divergence was defined between the distribution of the PF and a discrete approximation of the true  distribution computed on  a predefined grid. 
The ideas in \citet{fox2003adapting} were further explored by \citet{soto2005self}.  A heuristic approach based on the effective sample size was proposed by \citet{straka2006particle}, \cbrown{while \citet[Chapter 4]{Cornebise09thesis} pursued similar ideas with a scheme that regenerated particles until a certain performance criterion was met.} 
A disadvantage of using the effective sample size is that once a PF loses track of the hidden state, the effective sample size does not provide information for adjusting the number of particles. See other issues related to the effective sample size in \citet{elvira2018rethinking}.

\cbrown{A method for selecting the number of particles based on the particle approximation of the variance of the particle estimators was reported by \citet{lee2018variance}, where Feynman-Kac framework of \citet{DelMoral04} was used for the analysis. While well principled, this technique cannot be implemented online. \citet{Bhadra16} suggested an autoregressive model for the variance of the estimators produced by the PF was employed, but the resulting method works offline as well.} In a group of papers on alive PFs, the number of particles is adaptive and based on sampling schemes that ensure  a predefined number of particles to have non-zero weights \citep{LeGland05,Jasra13,DelMoral15alive}. In \citet{martino2017cooperative}, a fixed number of particles is adaptively allocated to several candidate models according to their performances. In \citet{Hu08}, particle sets of the same size are generated until an estimation criterion for their acceptance is met.

\subsection{\cbrown{Some background}} \label{ssIntroBackground}

\cbrown{In \citet{elvira2017adapting}, we introduced a methodology for assessing the convergence of PFs that works online and can be applied to a very broad class of state-space models and versions of the PF. The method is based on simulating fictitious observations from one-step-ahead predictive distributions approximated by the PF and comparing them with actual observations that are available at each time step. In the case of one-dimensional observations, a statistic is constructed that simply represents the number of fictitious observations which are smaller than the actual observation. It is proved in \citet{elvira2017adapting} that, as the PF converges, the predictive statistics become uniform on a discrete support and independent over time. From that realization, we proposed an algorithm for statistically testing the uniformity of the predictive statistic and, based on the test, update the number of particles in the PF.}  
\subsection{Contributions}

\cbrown{ 
In this paper, we propose a general block-adaptive PF where the number of particles is updated periodically, every $W$ discrete time steps. It is a rather general scheme that provides a common framework for the procedures described in \citet{elvira2017adapting} and enables us to introduce different versions of the algorithm and to extend the analysis of the methodology.} 

\cbrown{In particular, we first tackle the problem of whether the updates in the number of particles carried out at the end of each block of length $W$ translate into changes to the theoretical error bounds for the Monte Carlo estimators. While this is the kind of performance that one would \textit{like} to have (e.g., we want to see smaller errors when we increase the number of particles), what the standard arguments for proving the convergence of the PF\footnote{Either by induction as in \citet{Crisan01,Bain08} or using the contraction properties of Markov kernels as in \citet{DelMoral04,Kunsch05}.} yield directly are error bounds that depend on the minimum of the number of particles over time. Here, we use the approach of \citet{DelMoral04} to prove that, assuming that the state sequence is Markov and mixing, the approximation errors at the end of each block are bounded by the sum of two terms: one that depends on the number of particles in the current block and another one that decreases exponentially with the block length $W$.}    

\cbrown{
Next, we turn our attention to the analysis of the impact of the number of fictitious observations, $K$, used by the algorithm to compute the predictive statistics. We first prove that if the predictive statistics with $K$ fictitious observations are uniformly distributed, then the particle approximation of the filtering distribution match at least the first $K$ elements in a series of moments that characterize the true filter completely. Let us remark that this result is (close to) a converse to Theorem 2 in \citet{elvira2017adapting}: the latter says that when the PF converges the predictive statistics become uniform, while the new result says that if the predictive statistics with $K$ fictitious observations become uniform then the PF necessarily converges to match at least $K$ moments of the true filter. From this analysis, we deduce an alternative predictive statistic that can be computed (for some models) without sampling any fictitious observations at all and establish its connection with the original one.}

\cbrown{ 
Finally, we conduct an extensive simulation study that illustrates the theoretical results and provides further insights into the complexity, performance and behavior of the new class of algorithms. We show, for example, that choosing larger values of $K$ leads to more accurate filters with a higher computational cost, while a smaller $K$ yields a faster filter using less particles (but yielding rougher errors). We also illustrate how the approximation errors change with the number of particles (as predicted by the theory) or how the adaptive PF stabilizes around the same number of particles no matter the initial condition (i.e., whether started with many or few particles). Our simulations also show that the new predictive statistic (without fictitious observations) is effective but sensitive to prediction errors and hence it leads to higher computational loads.
}

\subsection{Organization of the paper}

\cbrown{In the next section, we briefly describe particle filtering as a sequential Monte Carlo methodology, then we introduce our notation and a general block-adaptive PF that updates the number of particles periodically and admits different implementations.
In Section \ref{sec_theory}, we present our convergence analysis of block-adaptive PFs. 
In Section \ref{sec_K}, we provide a detailed analysis of the number of generated fictitious particles and introduce a new predictive statistic that does not require generation of fictitious particles. 
In the last two sections, we present results of numerical experiments and our conclusions, respectively. 
}

\section{ {Background}}
\label{section_PF}

\subsection{{State-space models and particle filtering}}

We investigate Markov state-space models described by the triplet of probability distributions
\begin{eqnarray}
\X_0 &\sim& p(\x_0), \label{eqPrior}\\
\X_t &\sim& p(\x_t|\x_{t-1}), \label{eqState}\\
\Y_t &\sim& p(\y_t|\x_t), \label{eqLikelihood}
\end{eqnarray}
where  
\begin{itemize}
\item $t \in \mathbb{N}$ denotes discrete time; 
\item $\X_t$ is the system state at time $t$, i.e., a $d_x \times 1$-dimensional random vector taking values in a state space $\mX \subseteq \Real^{d_x}$,
\item $p(\x_0)$ is the a priori probability density function (pdf) of the state,
\item $p(\x_t|\x_{t-1})$ is the conditional density of $\X_t$ given $\X_{t-1}=\x_{t-1}$;
\item $\Y_t$ is a $d_y \times 1$-dimensional observation vector at time $t$, where $\Y_t\in\mY \subseteq \Real^{d_y}$ and is assumed conditionally independent of all the other observations given $\X_t$, 
\item $p(\y_t|\x_t)$ is the conditional pdf of $\Y_t$ given $\X_t=\x_t$. It is often referred to as the {\em likelihood} of $\x_t$, when it is viewed as a function of $\x_t$ for some fixed $\y_t$.
\end{itemize}

Based on the model \eqref{eqPrior}--\eqref{eqLikelihood}, we aim at estimating the sequence of posterior probability distributions 
$p(\x_t|\y_{1:t})$, $t=1, 2, \ldots$, recursively. Many schemes addressing this task rely on the decomposition
\begin{equation}
p(\x_t|\y_{1:t}) \propto p(\y_t|\x_t) \int p(\x_t|\x_{t-1}) p(\x_{t-1}|\y_{1:t-1}) d\x_{t-1},
\nonumber
\end{equation}
that relates the so-called filtering pdf at time $t$, $p(\x_t|\y_{1:t})$, to the filtering density at time $t-1$, $p(\x_{t-1}|\y_{1:t-1})$.

Let us denote the filtering and the predictive posterior probability measures as 
\begin{equation}
\pi_t(d\x_t):=p(\x_t|\y_{1:t})d\x_t, \quad \xi_t(d\x_t):=p(\x_t|\y_{1:t-1})d\x_t.
\end{equation}
The measure $\pi_t$ does not provide any further characterization of the probability distribution compared to the density $p(\x_t|\y_{1:t})$, however, Monte Carlo methods (including PFs) yield an approximation of $\pi_t$, rather than the pdf $p(\x_t|\y_{1:t})$. Another function that plays a central role in the methods investigated in this paper is the predictive pdf of the observations, $p(\y_t|\y_{1;t-1})$. We denote the associated probability measure as
\begin{equation}
\mu_t(d\y_t) := p(\y_t|\y_{1:t-1}) d\y_t.
\nonumber
\end{equation}
It is well known that the predictive pdf is instrumental for  model inference  \citep{Andrieu10,djuric2010assessment,Chopin12,crisan2018nested}.

The goal of particle filtering algorithms is to estimate sequentially the probability measures $\{\pi_t\}_{t\ge 1}$ as the observations $\{\y_t\}_{t\ge 1}$ are collected. The basic method for accomplishing this is known as the bootstrap filter (BF) introduced by \citet{Gordon93} (see also \citet{Doucet00}). 

At time $t=0$, the algorithm applies standard Monte Carlo to approximate the prior probability distribution, i.e., we generate $M$ i.i.d. samples $\x_0^{(m)}$, $m=1, \ldots, M$, from the pdf $p(\x_0)$. The samples $\x_0^{(m)}$ are often termed {\em particles}. Assume that the particles can be propagated over time, in such a way that we obtain a Monte Carlo approximation of the filtering distribution at time $t-1$ given by the particle set $\{ \x_{t-1}^{(m)} \}_{m=1}^M$. At time $t$, the BF generates an estimate of $\pi_t$ recursively, by taking three steps:
\begin{enumerate}
\item Draw new particles $\bar\x_{t}^{(m)}$, $m=1, \ldots, M$, from the conditional pdf's, $p(\x_t|\x_{t-1}^{(m)})$. Note at at this step we essentially simulate the model dynamics by propagating the particles one step forward.
\item Compute normalized importance weights of the particles, denoted
$$
w_t^{(m)} \propto p(\y_t|\bar \x_t^{(m)}), \quad m=1, \ldots, M.
$$
These weights are proportional to the likelihood and they satisfy $\sum_{m=1}^M w_t^{(m)} = 1$.
\item Resample the particles $M$ times with replacement using the weights $\{ w_t^{(m)} \}_{m=1}^M$ as probabilities  \citep{Li15}. This yields the new (and non-weighted) particle set $\{ \x_t^{(m)} \}_{m=1}^M$.
\end{enumerate}

From the particles and their weights one can compute estimates of several probability measures and pdfs. The filtering measure $\pi_t$ can be approximated as
\begin{equation}
{
\pi_t^M(d\x) = 
\sum_{m=1}^M
w_t^{(m)} \delta_{\bar{\x}_t^{(m)}}(d\x),
\nonumber
}
\end{equation}
where $\delta_{{\bar{\x}}_t^{(m)}}$ represents the Dirac delta measure located at $
{\bar{\x}}_t^{(m)} \in \mX$.
Moreover, at time $t$, once $\Y_{1:t-1}=\y_{1:t-1}$ are available but $\Y_t=\y_t$ has not been observed yet, the predictive pdfs of $\X_t$, denoted $\tilde p_t(\x_t) := p(\x_t|\y_{1:t-1})$, and $\Y_t$, denoted $p_t(\y_t) := p(\y_t|\y_{1:t-1})$,  can be approximated as
\begin{eqnarray}
\tilde p_t^M(\x_t) &:=& 
{\frac{1}{M}\sum_{m=1}^M  p(\x_t|\x_{t-1}^{(m)})}, \quad \x_t \in \mX, \quad \text{and} \\
p_t^M(\y_t) &:=& \frac{1}{M} \sum_{m=1}^M p(\y_t| \bar \x_t^{(m)}),\quad \y_t \in \mY.
\label{eq_pred_obs}
\end{eqnarray}

A key parameter in the standard BF is the number of particles $M$, which determines both the computational cost of the algorithm and also the accuracy of any estimators computed using the particles and weights \citep{DelMoral04,Bain08}. While $M$ is fixed in conventional particle filtering methods, the focus of this paper is on algorithms where $M$ can be updated sequentially \citep{elvira2017adapting}. 

\begin{table}[!t]
  \centering
  \caption{General BF with block-adaptive number of particles, $M_n$.}  
  \begin{tabular}{|p{0.95\columnwidth}|}
    \hline
    \footnotesize

\begin{enumerate}
\item {\bf [Initialization]}
\begin{enumerate}
\item Draw independent samples $\x_0^{(m)}$ from the prior $p(\x_0)$ and assign uniform weights, i.e.,  
\begin{eqnarray}
\x_0^{(m)} &\sim& p(\x_0),\quad\quad m=1,\ldots,M_0, \quad \text{and}\nonumber\\
w_0^{(m)} &=& \frac{1}{M_0},\quad\quad m=1,\ldots, M_0. \nonumber
\end{eqnarray}

\item Set $n=0$ (block counter) and choose $W_0>0$ (size of the first block).
\end{enumerate}

\item {\bf [For $t=1, 2, ...$]}
\begin{enumerate}

\item {\bf Bootstrap particle filter:}
\begin{itemize}[label={--}]
\item Sample $\bar{\x}_t^{(m)} \sim p(\x_t|\x_{t-1}^{(m)})$, $m=1,\ldots,M_n$.%
\item Compute normalized weights, $\bar{w}_{t}^{(m)} \propto p(y_t|\bar{\x}_t^{(m)})$, $m=1, \ldots, M_n$. 
\end{itemize}
\item {\bf Fictitious observations:}
\begin{itemize}[label={--}]
\item Draw $\tilde{y}_t^{(k)} \sim  p_t^M(y_t), \quad k=1, \ldots, K$. 
\item Compute $a_{K,M_n,t}$, i.e., the position of the actual observation $y_t$ within the set of ordered fictitious observations $\{ \tilde{y}_t^{(k)} \}_{k=1}^{K}$.
\end{itemize}
\item \textbf{Assessment of convergence:} If $t=\sum_{j=0}^n W_j -1$ (end of the $n$-th block) then:
\begin{itemize}[label={--}]
\item Analyze the subsequence
$$
\mathcal{S}_n = \{a_{K,M_n,t}, a_{K,M_n,t-1},\ldots,a_{K,M_n,t-W_n+1}\}
$$
with some specific algorithm from Section \ref{sec_previous_algs}.
\item Set $n=n+1$.
\item Select the number of particles $M_n>0$.
\item Select the block size $W_n>0$.
\item Resample $M_n$ particles with replacement, from the weighted set $\{ \bar{\x}_t^{(m)}, w_t^{(m)} \}_{m=1}^{M_{n-1}}$, to obtain $\{ \x_t^{(m)} \}_{m=1}^{M_n}$.
\end{itemize}
Else:
\begin{itemize}[label={--}]
\item Resample $M_n$ particles with replacement, from the weighted set $\{ \bar{\x}_t^{(m)}, w_t^{(m)} \}_{m=1}^{M_n}$, to obtain $\{ \x_t^{(m)} \}_{m=1}^{M_n}$.
\end{itemize}
\end{enumerate}

\end{enumerate} \\
    \hline
\end{tabular}
\label{tAdaptiveBF}
\end{table}

\subsection{{Block-adaptive selection of the number of particles}}

A generic block-adaptive method for selecting the number of particles is summarized in Table \ref{tAdaptiveBF}. Hereafter, we assume that the observations are one-dimensional (and hence we denote them as $y_t$ instead of $\y_t$) unless explicitly indicated. The methods to be described can be adapted to systems with multidimensional observations in a number of ways -- see \citet[Section IV-E]{elvira2017adapting} for a discussion on this topic.
Also note that we implement the algorithm based on the standard BF, but it is straightforward to extend the methodology to other PFs.

The block-adaptive method proceeds as follows. In Step 1(a) of Table \ref{tAdaptiveBF}, the filter is initialized with $M_0$ particles.
The particle filter works at each time step in a standard manner with the current number of particles, as described in Step 2(a). The first modification with respect to (w.r.t.) the BF comes in Step 2(b), where $K$ fictitious observations $\{\tilde{y}_t^{(k)}\}_{k=1}^K$ are simulated from the (approximate) predictive distribution of the observations $p^M_t(y_t)$, 
(see \citet[Section IV-A]{elvira2017adapting} for additional details). These fictitious observations are used to evaluate the statistics $A_{K,M,t} = a_{K,M,t}$, where $a_{K,M,t}$ is the position of the actual observation $y_t$ within the set of ordered fictitious observations $\{ \tilde{y}_t^{(k)} \}_{k=1}^{K}$.

The algorithm works {with} windows of varying size $W_n$, where at the end of the $n$th window (Step 2(c)), the sequence 
\begin{align}
\mathcal{S}_n =\{&a_{K,M_n,t-W_n+1}, a_{K,M_n,t-W_n+2}, \ldots \notag\\
&..., a_{K,M_n,t-2}, a_{K,M_n,t-1}, a_{K,M_n,t}\}
\label{eqSubSeqS}
\end{align} 
is processed for assessing the convergence of the filter. The number of particles is adapted (increased, decreased, or kept constant) based on the assessment.
When we assume that
\begin{itemize}
\item the fictitious observations $\{ \tilde y_t^{(k)} \}_{k=1}^K$ are independently drawn from the same pdf as the actual  observation $y_t$, and 
\item the statistic $A_{K,M,t}$ becomes independent of $M$, i.e., $A_{K,M,t} = A_{K,t}$ is exact as well, 
\end{itemize}
it is relatively straightforward to prove  the two propositions below \citep{elvira2017adapting}.

\begin{Proposicion} \label{prop_pmf}
If $y_t, \tilde{y}_t^{(1)}, \ldots, \tilde y_t^{(K)}$ are i.i.d. samples from a common continuous (but otherwise arbitrary) probability distribution, then the pmf of the random variable (r.v.) $A_{K,t}$ is
\beq
\label{eq_pmf_uniform}
\mbbQ_{K}(n) = \frac{1}{K+1}, \qquad n=0,...,K.
\eeq
\end{Proposicion}

\begin{Proposicion} \label{prop_indep_old}
If the r.v.'s $y_t, \tilde{y}_t^{(1)}, \ldots, \tilde y_t^{(K)}$ are i.i.d. with common pdf $p_t(y)$, then the r.v.'s in the sequence $\{ A_{K,t} \}_{t \ge 1}$ are independent.
\end{Proposicion}

In practical terms, Propositions \ref{prop_pmf} and \ref{prop_indep_old} suggest that when the approximation errors in the PF are small, i.e., $p_t^M(dy_t) \approx p_t(dy_t)$, we can expect the statistics in the sequence $\mathcal{S}_n$ of Eq. \eqref{eqSubSeqS} to be (nearly) independent and uniformly distributed. Therefore, testing whether the variates
$$
a_{K,M_n,t-W_n+1}, a_{K,M_n,t-W_n+2}, ..., a_{K,M_n,t-1},a_{K,M_n,t},
$$
are independent and/or uniform is an indirect manner of assessing the convergence of the PF. The key advantage of this approach is that Propositions \ref{prop_pmf} and \ref{prop_indep_old} do not depend on the specific choice of the transition density $p(\x_t|\x_{t-1})$ and the likelihood $p(y_t|\x_t)$, and therefore the statistics can be applied to a very general class of state-space models. A detailed analysis of the approximation errors in the statistic $A_{K,M,t}$ and its pmf $\mbbQ_{K,M,t}(n)$ is provided in \citet{elvira2017adapting}. In particular, it is proved that $\lim_{M\rw\infty} \mbbQ_{K,M,t} = \mbbQ_{K}(n)$ almost surely (a.s.) for every $t$.

\subsection{{Algorithms for adapting the number of particles}}
\label{sec_previous_algs}
 
We outline two specific techniques that assess the statistics $A_{K,M,t}$,. They exploit the properties of uniform distribution and statistical correlation.  

\subsubsection{\textbf{Algorithm 1.} Uniformity of $A_{K,M,t}$}
\label{sec_alg_unif}
Under the null hypothesis of perfect convergence of the PF (i.e., $p_t^M(dy_t) = p_t(dy_t)$), the r.v.'s $A_{K,M,t}$ are statistically independent and uniform. Therefore, we test if the variates in the subset $\mathcal{S}_n$ are i.i.d. uniform draws from the set $\{0, \ldots, K\}$. This is the scheme originally proposed in \citet{elvira2017adapting}, and it exploits Proposition \ref{prop_pmf} and Proposition \ref{prop_indep_old}. 

\subsubsection{\textbf{Algorithm 2.} Correlation of $A_{K.M,t}$}
\label{sec_alg_corr}
Under the same null hypothesis of perfect convergence of the PF, the variates in $\mathcal{S}_n$ are i.i.d. Since independence implies absence of correlation, we can test if the samples of $\mathcal{S}_t$ are correlated, e.g., using the scheme in \citet{elvira2016novel}. 
Note that in estimating the autocorrelation of $A_{K,M,t}$, longer windows (larger values of $W_n$) may be needed to improve accuracy. However, larger block-sizes imply a loss of responsiveness in the adaptation of $M$. 

%
\section{Error bounds for block-adaptive particle filters}
\label{sec_theory}

We present an analysis of the class of block-adaptive filters outlined in Table \ref{tAdaptiveBF}, with either fixed ($W_n=W$ for all $n$) or adaptive ($W_n$ updated together with $M_n$) block size from a viewpoint that was ignored in \citet{elvira2017adapting}. To be specific, we prove that at the end of the $n$th window,\footnote{Specifically, at time $t_n = \sum_{j=0}^n W_j - 1$.} the error bounds for the estimators that are computed using the {weighted particle set $\{ w_t^{(m)}, \bar \x_{t_n}^{(m)} \}_{m=1}^{M_n}$} can be written as a function of the {\em current} number of particles $M_n$ -- provided that the optimal filter $\pi_t$ satisfies a stability condition \citep{DelMoral04}. If one were to rely directly on classical convergence results for algorithms with fixed $M_n=M$ (see, e.g., \citet{DelMoral01c,Crisan02,DelMoral04,Kunsch05,Miguez13b}), the error bound at time $t_n$ would be characterized as a function of the minimum of the number of particles employed up to that time, namely 
$$
M_n^{\text{min}} := \min_{0 \le j \le n} M_j.
$$ 
The current number of particles, $M_n$, can be considerably larger than $M_n^{\text{min}}$ and, as a consequence, the error bound can be remarkably smaller.

%
\subsection{Notation}

{
For notational clarity and conciseness, let
\begin{equation}
\kappa_t(d\x_t|\x_{t-1}) := p(\x_t|\x_{t-1})d\x_{t}
\nn 
\end{equation} 
denote the Markov kernel that governs the dynamics of the state sequence $\{ \x_t \}_{t>0}$} and write
\begin{equation}
g_t^{y_t}(\x_t) := p(y_t|\x_t)
\nn
\end{equation}
to indicate the conditional pdf of the observations.

We analyze the algorithm outlined in Table \ref{tAdaptiveBF}, which is essentially a BF with $M_n$ particles in the $n$th time window, $\sum_{j=0}^{n-1} W_j \le t < \sum_{j=1}^n W_j$, where $W_j$ is the length of the $j$th window. As pointed out, the theoretical results we introduce are valid both for variable window lengths as well as for fixed $W_n=W$. Our analysis also holds independently of the update rule for $M_n$, as long as only positive values are permitted. Specifically, we assume that there is a positive lower bound $\underline M$ such that $M_n \ge \underline M$ for every $n \ge 0$. In practice, we usually have a finite upper bound $\overline M \ge M_n$ as well (but this plays no role in the analysis).

For any integrable real function $f : \mX \rw \Real$ and a probability measure $\alpha$, we use the shorthand notation
$$
(f,\alpha) = \int f(\x) \alpha(d\x)
$$
for integrals with respect to $\alpha$. If $\alpha$ has a pdf $a(\x)$, we also denote
$
(f,a) := (f,\alpha)
$
when convenient. Intuitively, we aim at proving that the bounds for the approximation errors $|(f,\pi_t^{M_n})-(f,\pi_t)|$, where
$$
{
(f,\pi_t^{M_n}) = \int f(\x) \pi_t^{M_n}(d\x) = \sum_{m=1}^{M_n} w_t^{(m)} f(\bar \x_t^{(m)}),
}
$$
effectively change when the number of particles $M_n$ is updated. Since the measures $\pi_t^{M_n}$ are random, the approximation errors $(f,\pi_t^{M_n})-(f,\pi_t)$ are real r.v.'s, and we can assess their $L_p$ norms. We recall that for a real r.v. $Z$ with probability measure $\alpha$, the $L_p$ norm of $Z$, with $p \ge 1$, is 
\begin{equation}
\| Z \|_p := \left( \mbbE\left[ |Z|^p \right] \right)^{\frac{1}{p}} = \left( \int |z|^p \alpha(dz) \right)^{\frac{1}{p}},
\nonumber
\end{equation}
where $\mathbb{E}[\cdot]$ denotes the expected value w.r.t. the distribution of the r.v. 

%
\subsection{Error bounds}
\label{sec_varying_N}

We show hereafter that by the end of the $n$th block of observations, the approximation  error
\beq
\| (f,\pi_{t_n}^{M_n}) - (f,\pi_{t_n}) \|_p, \quad \text{where $t_n = \sum_{j=0}^n W_j - 1$},
\nn
\eeq
and $f$ is a bounded real function, can be upper bounded by a function that depends on the current number of particles $M_n$ \cbrown{and ``forgets'' past errors exponentially fast.} This is true under certain regularity assumptions that we detail below.

Let us introduce the prediction-update (PU) operators $\Psi_t$ that generate the sequence of filtering probability measures $\pi_t$ given a prior measure $\pi_0$, the sequence of kernels $\kappa_t$ and the likelihoods $g_t^{y_t}$. 
\begin{Definition} \label{defPU}
Let $\mB(\mX)$ denote the Borel $\sigma$-algebra of subsets of $\mX$ and let $\mP(\mX)$ be the set of probability measures on the space $(\mX,\mB(\mX))$. We construct the sequence of PU operators $\Psi_t : \mathcal{P}(\mX) \rw \mP(\mX)$, $t \ge 1$, that satisfy
\begin{equation}
\left( f, \Psi_t(\alpha) \right) = \frac{ 
  \left(
    fg_t^{y_t},\kappa_t\alpha 
  \right)
}{
  (g_t,\kappa_t\alpha)
}, \quad t=1, 2, ...,
\label{eqDefPU}
\end{equation}
for any $\alpha \in \mP(\mX)$ and any integrable real function $f$, 
{where $\kappa_t\alpha(d\x_t)=\int \kappa_t(d\x_t|\x')\alpha(d\x')$ is the result of applying the Markov kernel $\kappa_t$ to the probability measure $\alpha$.}\footnote{
{Note that if $\alpha=\pi_{t-1}$, then $\kappa_t\pi_{t-1}(d\x_t) = p(\x_t|\y_{1:t-1})d\x_t$.}
}
\end{Definition}
It is not hard to see that Definition \ref{defPU} implies that $\pi_t = \Psi_t(\pi_{t-1})$. In order to represent the evolution of the sequence of filtering measures over several time steps, we introduce the composition of operators
\begin{equation}
\Psi_{t|t-r}(\alpha) := \left( \Psi_t \circ \Psi_{t-1} \circ \cdots \circ \Psi_{t-r+1} \right)(\alpha).
\end{equation} 
It is apparent that $\pi_t = \Psi_{t|t-r}(\pi_{t-r})$. The composition operator $\Psi_{t|t-r}$ is most useful for representing the filters obtained after $r$ consecutive steps when we start from different probability measures at time $t-r$, i.e., for comparing $\Psi_{t|t-r}(\alpha)$ and $\Psi_{t|t-r}(\beta)$ for $\alpha, \beta \in \mP(\mX)$.

{
In our analysis, we assume that the kernels $\kappa_t(d\x_t|\x_{t-1})$ satisfy a \textit{mixing} assumption \citep{DelMoral04,Kunsch05}. While this can be stated in various ways, we follow the approach in \cite{DelMoral04}, which relies on the composition of kernels to {mix} sufficiently over several time steps.
}

\begin{Assumption}[Mixing kernel]
\label{asMixing}
Let us write
\beq
\kappa_{t|t-m}=\kappa_t \circ \kappa_{t-1} \circ \cdots \circ \kappa_{t-m+1}
\eeq 
for the composition of $m$ consecutive Markov kernels. For every $S \in \mB(\mX)$ and integer $m \ge 1$, there exists a constant $\varepsilon_m > 0$, independent of $t$ and $S$, such that
{
\beq
\inf_{\x_{t-m},\x_{t-m}^\prime \in \mX} \frac{
	\kappa_{t|t-m}(S|\x_{t-m})
}{
	\kappa_{t|t-m}(S|\x^{{\prime}}_{t-m})
} > \varepsilon_m.
\nn
\eeq
}
\end{Assumption}

{
Assumption \ref{asMixing} implies that the sequence of optimal filters generated by the operators $\Psi_t$, $t \ge 1$, is stable \citep{DelMoral01c}. To be specific, it can be proved \citep{DelMoral01c,DelMoral04} that
\beq
\lim_{r\rw\infty} \sup_{\alpha,\beta \in \mP(\mX)} \left| 
  \left(
    f,\Psi_{t|t-r}(\alpha)
  \right) - \left(
    f,\Psi_{t|t-r}(\beta)
  \right) 
\right| = 0
\nn
\eeq
exponentially fast. The intuitive meaning is that such sequences ``forget'' their initial condition over time. It also implies that approximation errors are also forgotten over time when propagated through the operators $\Psi_t$, a fact that is often exploited in the analysis of PFs. 
}

The strongest assumption in our analysis is that the sequence of likelihoods is uniformly bounded away from zero (as well as upper bounded), as specified below.
{
\begin{Assumption}[Bounds]
\label{asLikelihood}
There exists a constant $\gamma>0$ such that
\begin{equation}
0 < \gamma < g_t^{y_t}(\x) \le 1
\end{equation}
for every $t \ge 1$ and every $\x \in \mX$.
\end{Assumption}
}

Assumption \ref{asLikelihood} depends not only on the form of the likelihood $g_t^{y_t}(\x)=p(\y_t|\x_t)$ but also on the specific sequence of observations $y_1, y_2, \ldots$ While it may appear restrictive, this is rather typical in the analysis of PFs (see \citet{DelMoral04,Kunsch05,Gupta15,crisan2017uniform}) and is expected to hold naturally when the state space $\mX$ is compact (as well as in other typical scenarios\footnote{For example, suppose that the observations are collected by sensors with limited sensitivity. To see this, consider a sensor located at $\s$ that measures the power transmitted by an object located at $\x$. Assuming free space, the received power (in dB's) can be modeled as $y = 10\log_{10}\left( P_0\|\s-\x\|^{-2} + \eta \right) + z$, where $z \sim N(0,\sigma^2)$ is Gaussian noise, $P_0$ is the transmitted power, and the parameter $\eta>0$ determines the sensitivity of the sensor. The likelihood function is 
$$
g^y(\x) \propto \exp\left\{ 
	-\frac{1}{2\sigma^2} \left( y - 10\log_{10}\left( P_0 \|\s-\x\|^{-2} + \eta \right) \right)^2 
\right\}.
$$ 
As a consequence, when $\| \s-\x \|\rw\infty$ the sensor observes $y = 10\log_{10}(\eta) + z$ independently of the target position $\x$ and, in particular, for fixed $\s$ we have $\lim_{\|\x\| \rw \infty} g^y(\x) \propto \exp\left\{ 
	-\frac{1}{2\sigma^2} \left( y - 10\log_{10} \eta \right)^2 
\right\} > 0$. Intuitively, the sensor cannot ``see'' targets which are too far away.}). 
{Also note that any bounded likelihood function can be normalized to guarantee $g_t^{y_t}\le 1$.}

The error bounds for estimator $(f,\pi_{t_n}^{M_n})$ are made precise by the following statement.

{
\begin{Teorema} \label{thConvergence}
Let $t_n = \sum_{j=0}^n W_j-1$ and let $\pi_{t_n}^{M_n}$ be the particle approximation of the filtering measure $\pi_{t_n}$ produced by the block-adaptive BF in Table \ref{tAdaptiveBF}. If Assumption \ref{asMixing} (mixing kernel) and Assumption \ref{asLikelihood} (bounds) hold, then for any $p\ge 1$, 
\beqa
&\sup_{|f|\le 1} \left\|
  (f,\pi_{t_n}^{M_n}) - (f,\pi_{t_n})
\right\|_p < \frac{
	m C
}{
	\gamma^{2m-1} \varepsilon_m^3 \sqrt{M_n}
}& \nn\\
&+ \frac{
	2\left(
		1- \gamma^{m-1}\varepsilon_m^2
	\right)^{\lfloor \frac{W_n}{m} \rfloor} 
}{
	\gamma^m\varepsilon_m
}\sup_{|f|\le 1} \left\|
  	(f,\pi_{t_{n-1}}^{M_{n-1}}) - (f,\pi_{t_{n-1}})
\right\|_p,&
\nn
\eeqa
{where $\sup_{|f|<1}$ denotes the supremum over all real functions $f:\mX\mapsto\Real$ with $\|f\|_\infty\le 1$. The real constants  $C<\infty$, $\gamma>0$ and $\varepsilon_m>0$, as well as the integer $m \ge 1$, are independent of $n$, $M_n$ and $W_n$}. 
\end{Teorema}
}
See Appendix \ref{apConvergence} for a proof. 

{
The theorem states that $W_n$ can be chosen in such a way that the ``inherited error'' due to, e.g., a lower number of particles $M_{n-1}$ in the $(n-1)$th block can be forgotten when a sufficiently large window length $W_n$ is selected in the $n$th block. This is due to the stability property of the PU operator $\Psi_t$ (which is guaranteed under Assumption \ref{asMixing}). In particular, \beq
\lim_{W_n\rw\infty} \sup_{|f|\le 1} \left\|
  (f,\pi_{t_n}^{M_n}) - (f,\pi_{t_n})
\right\|_p < \frac{
	mC
}{
	\gamma^m \varepsilon_m^2 \sqrt{M_n}
}.
\nn
\eeq
}

%
\section{Analysis of the number of fictitious observations, $K$}
\label{sec_K}
The performance analysis of \citet{elvira2017adapting} establishes the main results needed for a principled, online adaptation of the number of particles $M$, but leaves a number of questions unanswered. One of them, whether the error bounds of the particle estimators change as we update the number of particles online, has been addressed in Section \ref{sec_theory}. Two other major questions are
\begin{itemize}
\item[(i)] whether the statistics $A_{K,M_n,t}$ becoming uniform is {\em sufficient} for the particle approximations $p_t^{M_n}$ and $\pi_t^{M_n}$ to converge towards $p_t$ and $\pi_t$, respectively (the analysis in \citet{elvira2017adapting} only shows that this is {\em necessary}), and  
\item[(ii)] how the choice of the number of fictitious observations $K$ affects the performance of the adaptive algorithm (i.e., the approximation error of either $p_t^{M_n}$ or $\pi_t^{M_n}$).
\end{itemize}

We tackle these two issues in this section. To be precise, we prove that if the statistics $A_{K,M_n,t}$ are uniform r.v.'s for every $K \in \mathbb{N}$, then the approximate predictive pdf $p_t^{M_n}(y)$ becomes equal to the actual density $p_t(y)$ almost everywhere. 
{This result serves as a converse for Theorem 2 in \cite{elvira2017adapting}} -- which states that $p_t^{M_n}(y) \stackrel{M_n\rw\infty}{\longrightarrow} p_t(y)$ implies that the statistics $A_{K,M_n,t}$ become uniform r.v.'s. Intuitively, the new result ensures that if the $A_{K,M,t}$'s are well-behaved then so are the particle approximations $p_t^{M_n}$. Our analysis also provides insight into the choice of $K<\infty$. Specifically, it yields a quantitative interpretation of how $p_t^{M_n}$ becomes closer to $p_t$ when $A_{K,M_n,t}$ is uniform for larger and larger $K$. 

As a by-product of this analysis, we identify an alternative statistic $B_t$ (and its particle estimator $B_{M_n,t}$) that can be used for assessing the performance of the particle filter without generating fictitious observations. This alternative statistic admits an interpretation as the limit of the sequence $K^{-1}A_{K,M_n,t}$ when ${K\rw\infty}$ and, therefore, it inherits the key theoretical properties of the statistics $A_{K,M_n,t}$.

\subsection{A converse theorem}
\label{sec_statistics}

Let us consider the true predictive density $p_t(y)$ and an approximation, computed via particle filtering or otherwise, that we denote as $\hat p_t(y)$. In this subsection, we drop the number of particles $M_n$ in the notation because the results to be presented are valid without regard for the type of approximation of the predictive distribution of the observations, that is, it is not important if it is approximated by a Monte Carlo-based method or is obtained via an analytical approach. The analysis in this section relies to a large extent on the properties of the cumulative distribution functions (cdf's) associated to $p_t(y)$ and $\hat p_t(y)$, which we denote as 
\begin{eqnarray}
F_t(a) &=& (\textbf{1}_{(-\infty,a)},p_t), \quad \text{and} \nonumber\\
\hat F_t(a) &=&  (\textbf{1}_{(-\infty,a)},{\hat p_t)}, \nonumber
\end{eqnarray}
respectively, where 
$$
\textbf{1}_A(y) = \left\{
	\begin{array}{ll}
		1, &\text{if $y\in A$}\\
		0, &\text{otherwise}\\
	\end{array}
\right.
$$
denotes the set-indicator function.

The statistic $\hat A_{K,t}$ is computed by generating $K$ i.i.d. fictitious observations 
{from} $\hat p_t$, denoted $y_t^{(1)}, $ $  \ldots, $ $ y_t^{(K)}$, and then computing the relative position of the actual observation $y_t$ (distributed according to $p_t$) within the ordered fictitious observations. From Proposition \ref{prop_pmf}, we know that if $p_t=\hat p_t$ then $\hat A_{K,t}$ is uniform for every $K \in \mathbb{N}$. Here, we pose the reverse question: if $\hat A_{K,t}$ is uniform for every $K \in \mathbb{N}$, can we claim that $p_t=\hat p_t$? Moreover, if only \textit{some} statistic $\hat A_{K,t}$ is uniform (i.e., for {\em some} finite $K \in \mathbb{N}$), can we expect $\hat p_t$ to be close to $p_t$ in some quantitative well-defined sense?

Our analysis relies on two basic results in probability theory.
\begin{Lema} \label{lemInversion}
Let $Y$ be a continuous real r.v. on a probability space $(\Omega, \mF, \Prob)$ and let $p$ denote a pdf, with cdf $F(\cdot)=( \textbf{1}_{(-\infty,\cdot)}, p )$. The r.v. $F(Y)$ has uniform distribution $\mathcal{U}(0,1)$ if, and only if, $Y$ is distributed according to $p$.  
\end{Lema}
\noindent \textit{\textbf{Proof:}} The inversion theorem 
(see, e.g., Theorem 2.1 in \cite{Martino18sampling})
guarantees that if $Y \sim p$ then the r.v. $F(Y)$ is $\mathcal{U}(0,1)$. For the reverse implication, assume $F(Y) \sim \mathcal{U}(0,1)$, which implies that $\Prob( F(Y) < a ) = a$ for any $a\in[0,1]$. As a consequence, 
$$
\Prob( Y < a ) = \Prob( F(Y) < F(a) ) = F(a),
$$ 
hence $Y \sim p$. \qed

\begin{Lema} 
\label{lemObvious}
Let $p$ denote a pdf with associated cdf $F(\cdot)=( \textbf{1}_{(-\infty,\cdot)}, p )$. For every $n \in \mathbb{N}$ we have $( F^{n}, p) = \frac{1}{n+1}$.
\end{Lema}
\noindent\textbf{\textit{Proof}}: Let $Y$ be a r.v. with pdf  $p$; from Lemma \ref{lemInversion} we have $F(Y) \sim \mathcal{U}(0,1)$, hence
$$
(F^n,p) = \mathbb{E}\left[ F(Y)^n \right] = \mathbb{E}[ U^n ],
$$
where $U \sim \mathcal{U}(0,1)$. It is straightforward to verify that $\mathbb{E}[ U^n ] = \frac{1}{n+1}$. \qed

Using the basic lemmas above, we establish the key result that relates the approximate cdf $\hat F_t$ to the true functions $F_t$ and $p_t$.

\begin{Teorema} \label{thMoments}
Assume the observation $Y_t$ is a continuous r.v. with a pdf $p_t$ and cdf $F_t$. Let the pdf $\hat p_t$ and its associated cdf $\hat F_t(\cdot)=(\textbf{1}_{(-\infty,\cdot)},\hat p_t)$ be estimates of $p_t$ and $F_t$, respectively. If the r.v. $\hat A_{K,t}$ 
{constructed from $\hat p_t$} is uniform then
\begin{eqnarray}
( \hat F_t^n, p_t ) = \frac{1}{n+1}, \;\; \forall n\in\{0, 1,...,K\}.
\label{eq_lemma_2}
\end{eqnarray}
\end{Teorema}
\noindent\textbf{\textit{Proof}}: See Appendix \ref{appendix_lemma2}. \qed

\begin{Nota} \label{remBt}
Let $Y_t$ be the actual observation with pdf $p_t$. Given the actual cdf $F_t$ and its estimate $\hat F_t$, we can construct the r.v.'s $B_t=F_t(Y_t)$ and $\hat B_t=\hat F_t(Y_t)$. From Lemma \ref{lemObvious}, we readily obtain 
$$
\mathbb{E}[ B_t^n ] = (F_t^n,p_t) = \frac{1}{n+1} \quad \mbox{for every $n \ge 0$.}
$$
However, Theorem \ref{thMoments} guarantees that if $\hat A_{K,t}$ is uniform, then 
$$
\mathbb{E}[ \hat B_t^n ] = (\hat F_t^n,p_t) = \frac{1}{n+1} \quad \mbox{for $n=0, \ldots, K$.}
$$
Therefore, if the statistic $\hat A_{K,t}$ is uniform, the r.v.'s $B_t=F_t(Y_t)$ and $\hat B_t = \hat F_t(Y_t)$ share their first $K$ moments. This is a quantitative characterisation of the similarity between $F_t$ and $\hat F_t$. In particular, if $\hat A_{K,t}$ is uniform for every $K \in \mathbb{N}$, we have $\mathbb{E}[ \hat B_t^n ] = \mathbb{E}[ B_t^n ] = \frac{1}{n+1}$ for every $n\in \mathbb{N}$ and, as a consequence, $\hat F_t=F_t$ and $\hat p_t(y) = p_t(y)$ almost everywhere in the observation space.
\end{Nota}
 
\subsection{Assessment without fictitious observations: the statistic $B_t$}

If $Y_t$ is a continuous r.v. with pdf $p_t$, then the sequence of statistics $B_t = F_t(Y_t)$, $t \in \mathbb{N}$, is i.i.d. with a common distribution $\mathcal{U}(0,1)$.\footnote{The fact that every $B_t$ is uniform is a consequence of Lemma \ref{lemInversion}. Independence can be proved by the same argument as in Proposition 3 of \citep{elvira2017adapting}.} From Remark \ref{remBt} it is apparent that we can use the particle filter to compute estimators $\hat B_t \equiv B_t^{M_n}$ over a window of observations (i.e., for $t_{n-1} < t \le t_n$) and then use the estimates to assess the performance of the filter. Two straightforward approaches to performing this assessment are:
\begin{itemize}
	\item testing for uniformity in (0,1) of the estimates $b_{t_{n-1}+1}^{M_n}, \ldots, b_{t_n}^{M_n}$ or
	\item evaluating the sample moments $\frac{1}{W_n} \sum_{t=t_{n-1}+1}^{t_n} \left( b_t^{M_n} \right)^m$, which should be close to $\frac{1}{m+1}$, according to Theorem \ref{thMoments}, when the particle filter is ``performing well.''
\end{itemize}

Since the approximate cdf of $Y_t$ computed via the particle filter $F_t^{M_n}(a) = ( \textbf{1}_{(-\infty,a)}, p_t^{M_n})$ is an integral w.r.t. $p_t^{M_n}$ and $B_t^{M_n} = F_t^{M_n}(Y_t)$, it follows that the estimates $B_t^{M_n}=b_t^{M_n}$ can be computed with $\mathcal{O}(M_n)$ operations, without generating any fictitious observations, as 
$$
b_t^{M_n} = ( \textbf{1}_{(-\infty,y_t)}, p_t^{M_n}) = \frac{1}{M_n} \sum_{m=1}^{M_n} \int_{-\infty}^{y_t} g_t^y(\bar \x_t^{m}) dy.
$$
Note, however, that calculating the $b_t^{M_n}$'s from the observation $y_t$ demands the ability to integrate the conditional pdf of the observations $g_t^y(\bar \x_t) = p(y|\bar \x_t)$. This is a straightforward numerical task when the observation noise is additive and Gaussian, but it may not be possible for other models.

Provided it can be computed, the statistic $B_t^{M_n}$ converges to the actual r.v. $B_t$ when $M_n \rw \infty$ under the basic assumptions in \citet{elvira2017adapting}, reproduced below for convenience (and restricted to the case of scalar observations).
\begin{itemize}
\item[\mfL] For each $t \ge 1$, the function $g_t$ is positive and bounded, i.e., $g_t^y(\x)>0$ for any $(y,\x) \in \mY \times \mX$ and $\| g_t \|_\infty = \sup_{(y,\x) \in \mY \times \mX} | g_t^y(\x) | < \infty$.

\item[\mfD] For each $t \ge 1$, the function $g_t^y(\x)$ is Lipschitz-continuous w.r.t. $y$.

\item[\mfC] For any $0 < \beta < 1$ and any $p \ge 4$, the sequence of intervals
\begin{equation}
C_M := \left[ -\frac{M^\frac{\beta}{p}}{2}, +\frac{M^\frac{\beta}{p}}{2} \right] \subset \Real
\nonumber
\end{equation}
satisfies the inequality $\mu_t(\overline{C_M}) \le b M^{-\eta}$ for some constants $b>0$ and $\eta>0$ independent of $M$ (yet possibly dependent on $\beta$ and $p$), where $\overline{C_M}=\Real\backslash C_M$ is the complement of $C_M$.
\end{itemize}
To be specific, we have the following result.

\begin{Proposicion} \label{propQM}
Let $Y_t$ be a r.v. with pdf $p_t(y_t)$, and let the observations $y_{1:t-1}$ be fixed. If the Assumptions \mfL, \mfD~and \mfC~hold, then there exists a sequence of non-negative r.v.'s $\{\varepsilon_t^{M_n}\}_{M_n \in \mathbb{N}}$ such that $\lim_{M_n\rw\infty} \varepsilon_t^{M_n} = 0$ a.s. and
\begin{equation}
B_{t} - \varepsilon_t^{M_n} \leq B_{M_n,t} \leq B_{t} + \varepsilon_t^{M_n}.
\label{eqGG}
\end{equation}
In particular, $\lim_{M_n\rw\infty} B_{M_n,t} = B_{t}$ a.s. and the distribution of $B_t^{M_n}$ converges to $\mathcal{U}(0,1)$ when $M_n\rw\infty$. 
\end{Proposicion}
\noindent \textit{\textbf{Proof}}: Recall that $B_t=( \textbf{1}_{(-\infty,Y_t)}, p_t)$ and $B_t^{M_n} = ( \textbf{1}_{(-\infty,Y_t)}, p_t^{M_n})$. Therefore, Proposition \ref{propQM} is a straightforward consequence of Theorem 1 in \citet{elvira2017adapting}, provided that Assumptions \mfL, \mfD, and \mfC \ hold. \qed

Finally, we note the strong connection between the statistics $B_t^{M_n}$ and $A_{K,M_n,t}$. Recall $A_{K,M_n,t}$ represents the number of fictitious observations that are smaller than $y_t$, while $B_t^{M_n}$ represents the probability $\int_{-\infty}^{Y_t} p_t^{M_n}(y)dy$. Intuitively, when $K\rightarrow\infty$, the empirical rate of observations smaller than $y_t$ should converge to the probability of a fictitious observation being smaller than $Y_t$. More precisely, we can state the proposition below.

\begin{Proposicion} \label{propA_B}
If $Y_t \sim p_t$ is a continuous r.v., then
\begin{equation}
\lim_{K\to\infty}\frac{A_{K,M_n,t}}{K} = B_t^{M_n}.
\nonumber
\end{equation}
\end{Proposicion}
\noindent \textit{\textbf{Proof}}: Recall that $B_t^{M_n} = ( \textbf{1}_{(-\infty,Y_t)}, p_t^{M_n} ) \le 1$. It is possible to estimate this integral by drawing $K$ samples $\tilde{y}_t^{(k)}$ from $\mu_t^M$ and building the standard Monte Carlo estimator $\frac{1}{K}\sum_{k=1}^K \textbf{1}_{(-\infty,y_t)}( \tilde{y}_t^{(k)}) = \frac{A_{K,M_n,t}}{K}$. Note that this estimator is unbiased, i.e., according to the strong law of large numbers,
\begin{equation}
\lim_{K\to\infty}\frac{A_{K,M_n,t}}{K}  \equiv \lim_{K\to\infty} \frac{1}{K}\sum_{k=1}^K \textbf{1}_{(-\infty,y_t)}( \tilde{y}_t^{(k)})  = B_t^{M_n}.
\nonumber
\end{equation}
\qed

\section{Numerical experiments}

In the first experiment, we show the relation between the correlation coefficient of 
$A_{K,M,t}$ and the MSE of an estimator obtained from the particle approximation in a non-linear {state-space model}. Then, we complement the results of Section \ref{sec_statistics}, showing numerically some properties of $A_{K,M,t}$ for different values of $K$ and $M$, and their connection to the statistic $B_{M,t}$. Third, we illustrate numerically the convergence of the block-adaptive BF.  
\begin{figure}
\centering
\subfigure[MSE in the estimate of the posterior mean.]{
\includegraphics[width=0.49\textwidth]{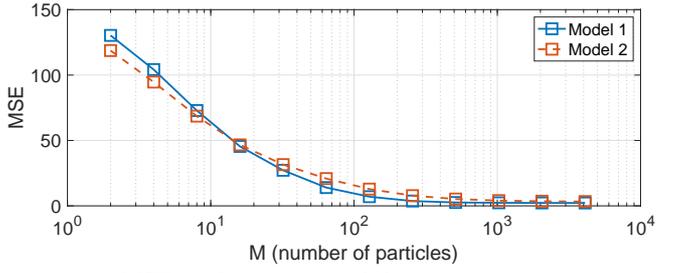}
} 
\centering
\subfigure[Algorithm 1 of Section~\ref{sec_previous_algs}. We show the p-value of the Pearson's $\chi^2$ test for assessing the uniformity of the statistic $A_{K,M,t}$.]{
\includegraphics[width=0.49\textwidth]{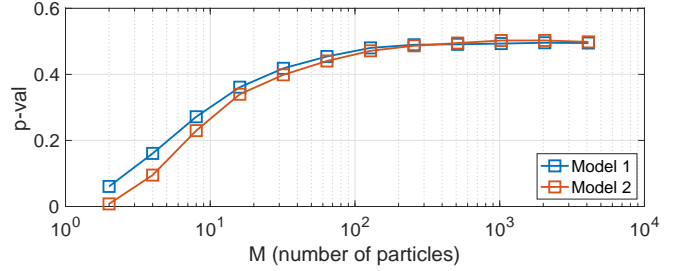}
} 
\centering
\subfigure[Algorithm 2 of Section~\ref{sec_previous_algs}. 
{The computed Pearson's correlation coefficient $r$ as a function of $M$.}  
]{
\includegraphics[width=0.49\textwidth]{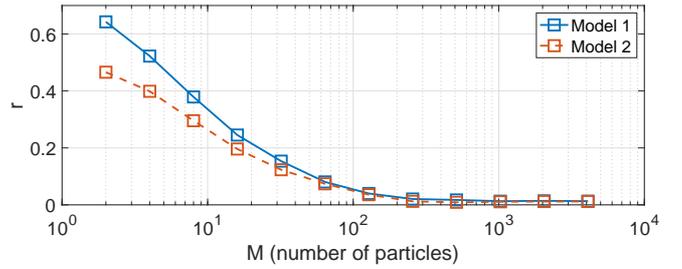}
} 
\caption{Stochastic growth model: MSE, p-value of the Pearson's $\chi^2$, and Pearson's correlation coefficient $r$.}
\label{sg_corr}
\end{figure}

\subsection{Assessing convergence from the correlation of 
$A_{K,M,t}$.}

Consider the stochastic growth model (see, e.g., \citet{djuric2010assessment}),
\beqa
x_t&=&\frac{x_{t-1}}{2} + \frac{25x_{t-1}}{1+x_{t-1}^2} + 8\cos(\phi t) +  u_t,\label{sgm_state}\\
y_t&=& \frac{x^2_t}{20}+ v_t,\label{sgm_obs}
\eeqa
where $\phi=0.4$, 
and $u_t$ and $v_t$ are independent Gaussian r.v.'s with zero mean, and variance $\sigma_u^2$ and $\sigma_v^2$, respectively. At this point, we define two models: 
\begin{itemize}
    \item Model 1: $\sigma_u = 1$ and $\sigma_v = 0.5$,
  \item Model 2: $\sigma_u = 2$ and $\sigma_v = 0.1$.
\end{itemize}
In this example, we ran the BF for $T=5,000$ time steps, always with a fixed number of particles $M$. We tested different values, namely 
$M \in \{2,2^2,2^3,...,2^{12}\}$. In order to assess the behavior of 
$A_{K,M,t}$, we set $K=7$ fictitious observations. 

Figure \ref{sg_corr}(a) shows the mean squared error (MSE) of the estimate of the posterior mean for each value of $M$, which obviously decreases as $M$ increases. Figure \ref{sg_corr}(b) {displays} 
{the p-value of the Pearson's $\chi^2$ test for assessing} 
the uniformity of $A_{K,M,t}$ (in the domain $A_{K,M,t} \in \{0,...,K+1 \}$) in windows of length $W=20$ (Algorithm 1 of Section \ref{sec_previous_algs}; see more details in \citet{elvira2017adapting}. 
Clearly, increasing the number of particles also increases the p-value, i.e., the distribution of the statistic becomes closer to the uniform distribution. Figure \ref{sg_corr}(c) is related to Algorithm 2 of Section~\ref{sec_alg_corr}. We show the sample Pearson's correlation coefficient $r$, using the whole sequence of statistics $\{ a_{K,M,t}\}_{t=1}^T$, computed 
{with a lag $\tau = 1$}. {All results are averaged over $200$ independent runs.}

We observe that when we increase $M$, the correlation between consecutive statistics decreases. 
It is interesting to note that the curve of the correlation coefficient $r$ has a very similar shape to the MSE curve and, hence, can be used as a proxy. While $r$ can be easily computed, the MSE is always unknown. This shows the utility of Algorithm~2.

It can be seen that both algorithms can  identify a malfunctioning of the filter when the number of particles is insufficient. We note that Algorithm 2 works better for Model 1 than for Model 2 because the autocorrelation of the statistics is more sensitive for detecting the malfunctioning  for low $M$. 
However, Algorithm 1 works better for Model 2 because the p-value of the uniformity test is always smaller than in Model 1, i.e., it is more discriminative. Therefore, there is no clear superiority of one algorithm over the other.

\subsection{Effect of the choice of the number of fictitious observations $K$}
\label{K_influence}
In this experiment, we evaluate the effect of the value $K$ in the performance of the uniformity test of the statistic $A_{K,M,t}$. We use the same model parameters as in the previous example. First, we fix the number of particles $M=\{2, 4, 16, 64, 256, 1024, 4096 \}$ for each run during the $T$ time steps (i.e., no adaptation is performed). Then, with $W=15$ and {$K\in\{ 2  ,         3    ,       5  ,         7 ,         10,          20 \}$}, we compute the p-value of the Pearson's $\chi^2$ test for assessing the uniformity of the statistic $A_{K,M,t}$. In Table \ref{table_pval_fixed_M}, we show the average of the p-value over $1,000$ independent runs for all combinations of $K$ and $M$. We can see that the misbehavior of the filter with a low number of particles $M$ can be detected regardless the number of fictitious observations $K$. For a larger $K$, in general the p-value decreases but it does not make a significant difference, which confirms our hypotheses in Section \ref{sec_K}: (a) the framework is robust to the selection of $K$; (b) increasing $K$ increases the detection power of the algorithms; (c) for reasonable values of $K$, when the filter misbehaves, the assessment of the uniformity of $A_{K,M,t}$ detects the misbehavior (and when the filter works well, there are not false alarms with large $K$); and (d) a small $K$ can be selected, which implies a low extra computational complexity of the proposed methodology.

\begin{figure}
\centering
\includegraphics[width=0.49\textwidth]{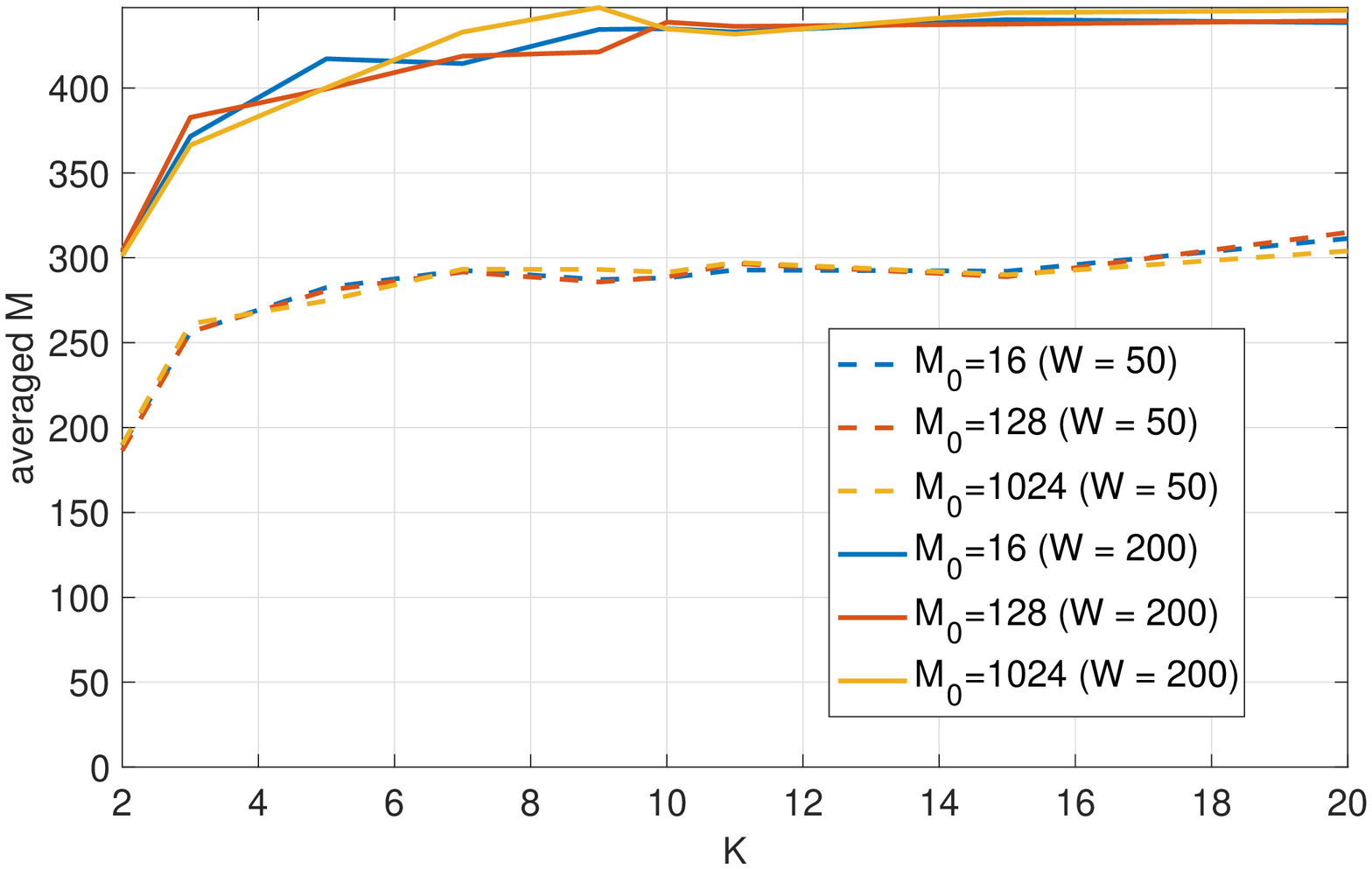}
\label{fig_averaged_M}
\caption{Number of particles averaged over the last {$50$} windows of the adaptive algorithm implemented as a function of the number fictitious observations {$K\in\{ 2  ,         3    ,       5  ,         7 ,         10,          20,\}$. Each curve represents different values of the window size $W\in\{50,200\}$ and initial number of particles $M_0 = \{16,128,1024\}$.}
}
\end{figure}

In a second experiment, we implement Algorithm~1 described in Section \ref{sec_previous_algs} on the same model, now using $T =10^4$ as the length of the time series. We set the algorithms parameters as $p_\ell = 0.2$, $p_h = 0.6$, $W\in\{50,200\}$, and an initial number of particles $M_0 = \{16,1024\}$. In Table \ref{table_averaged_M}, we show the resulting number of particles averaged over the last {$50$} windows of the adaptive algorithm implemented for each value of fictitious observations, {$K\in\{ 2  ,         3    ,       5  ,         7 ,         9, 10, 11, 15,         20 \}$}. The results are also averaged over $100$ independent runs. Figure \ref{fig_averaged_M} shows the same results of the averaged number of particles as a function of $K$. We see again that the selected number of particles does not depend much on $K$, as observed in the previous experiment. We also see that the window length does have an effect,  requiring a higher number of particles when $W$ is larger. The reason is that a larger $W$ implies that more realizations of the statistic are observed, so in cases where the filter is tracking but with some non-negligible errors, it is more likely that the statistical test rejects the null hypothesis whenever more evidence is accumulated. 

\begin{figure}
\centering
\subfigure[$M_0=1024$]{
\includegraphics[width=0.49\textwidth]{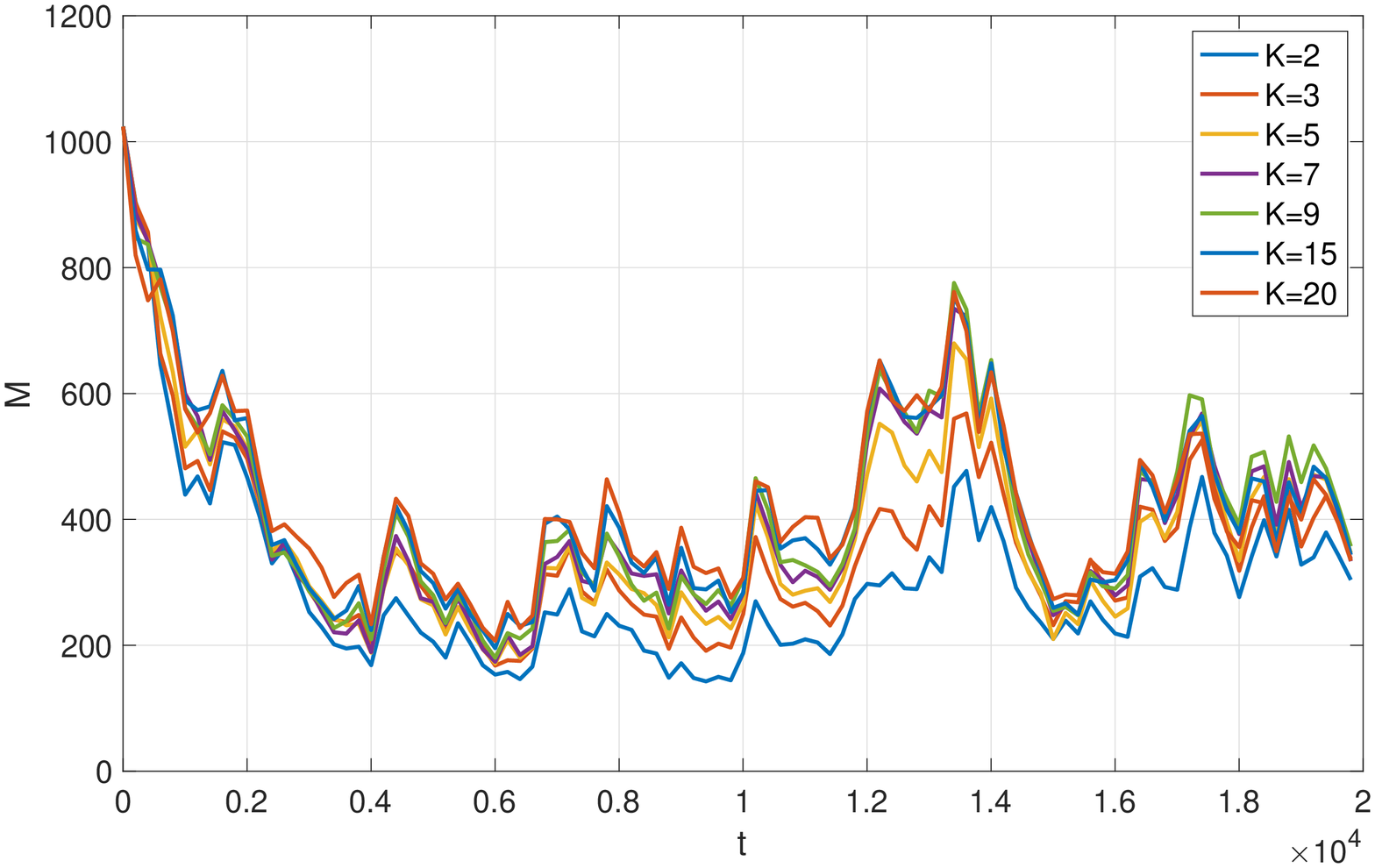}
} 
\centering
\subfigure[$M_0=128$]{
\includegraphics[width=0.49\textwidth]{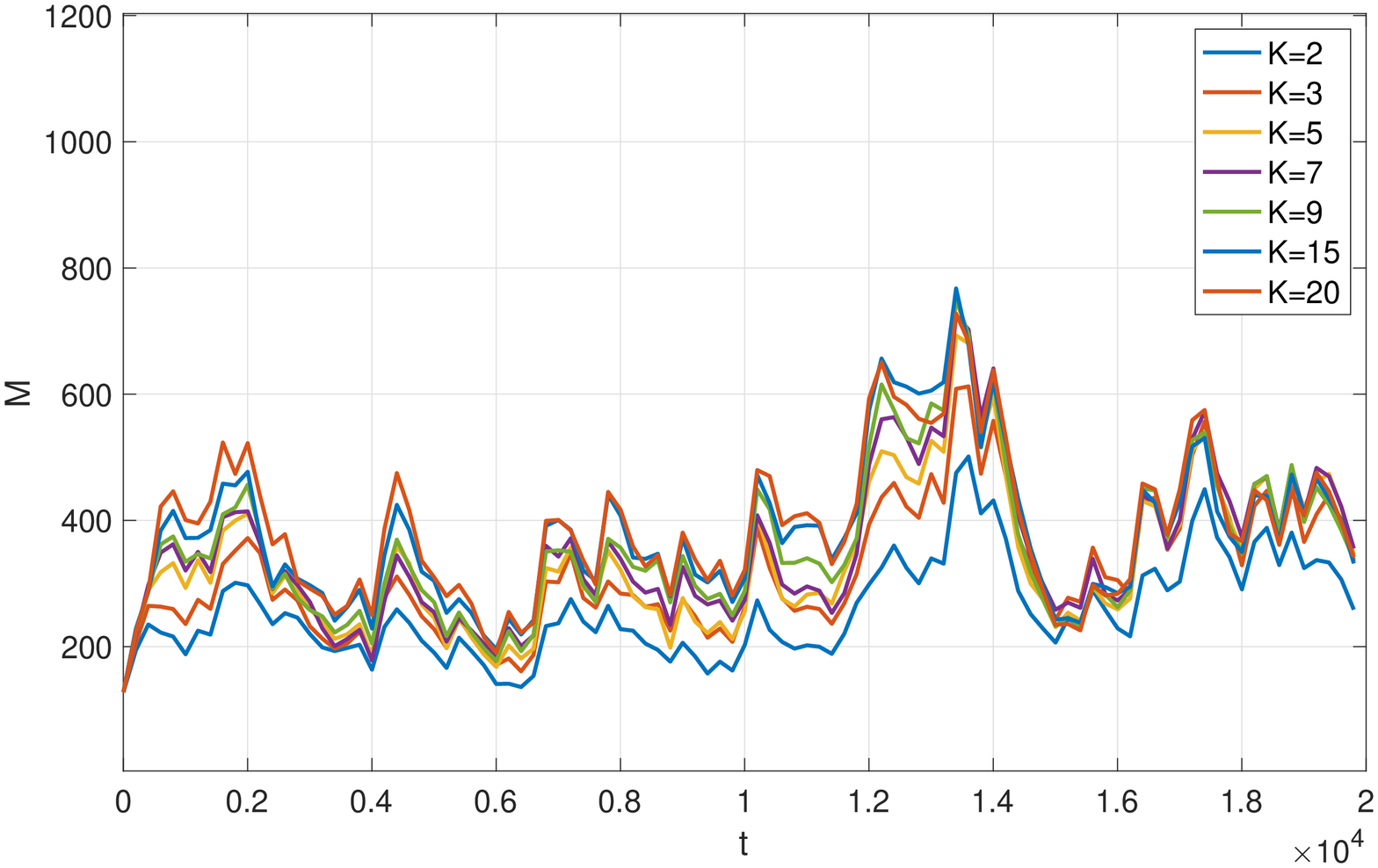}
} 
\centering
\subfigure[$M_0=16$]{
\includegraphics[width=0.49\textwidth]{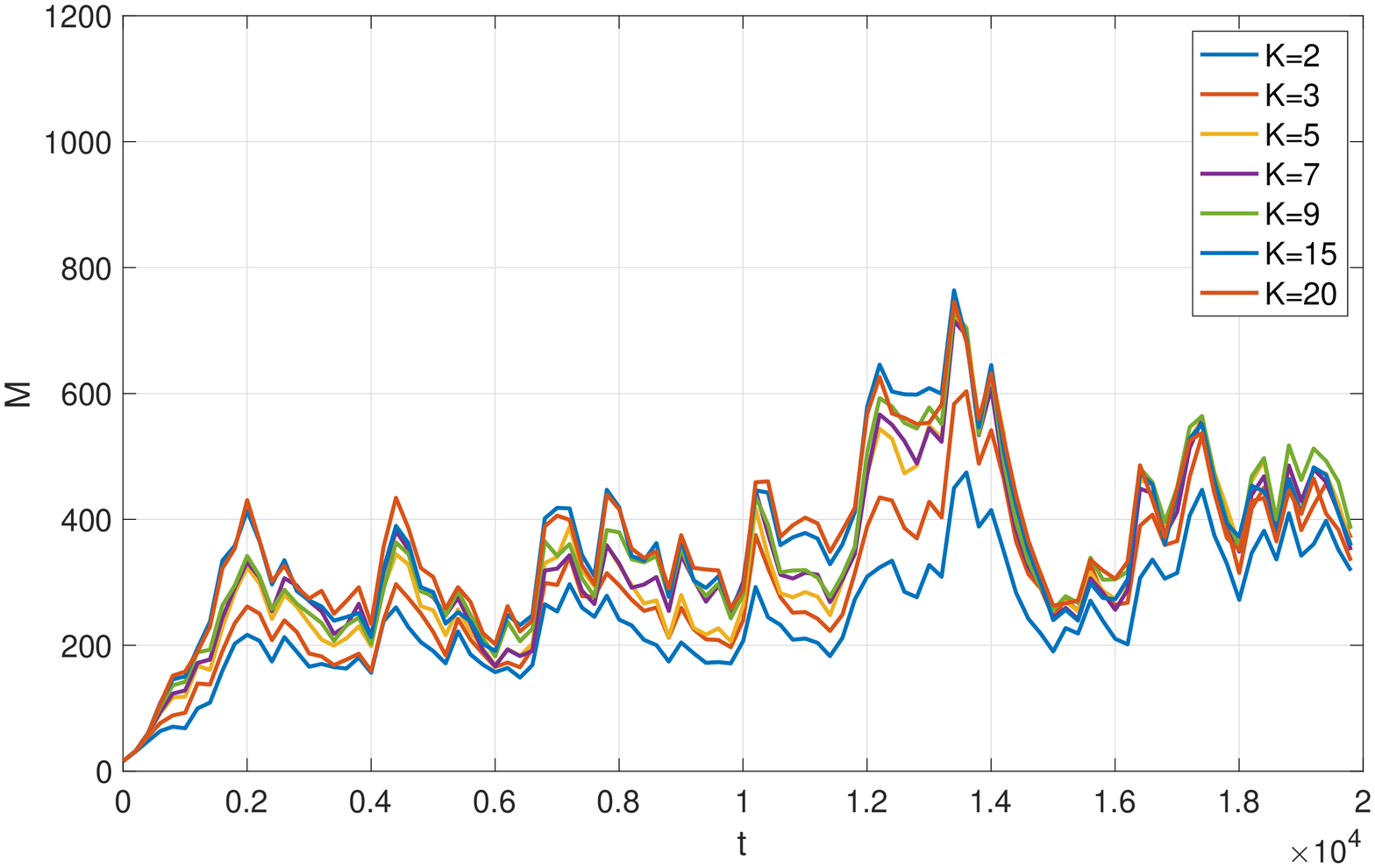}
} 
\caption{Averaged number of particles as a function of time for a fixed number of initial number of particles $M_0 = \{16,128,1024\}$.}
\label{fig_adaptive_M_vs_time_fixed_M0}
\end{figure}

\begin{figure}
\centering 
\subfigure[$K=5$]{
\includegraphics[width=0.49\textwidth]{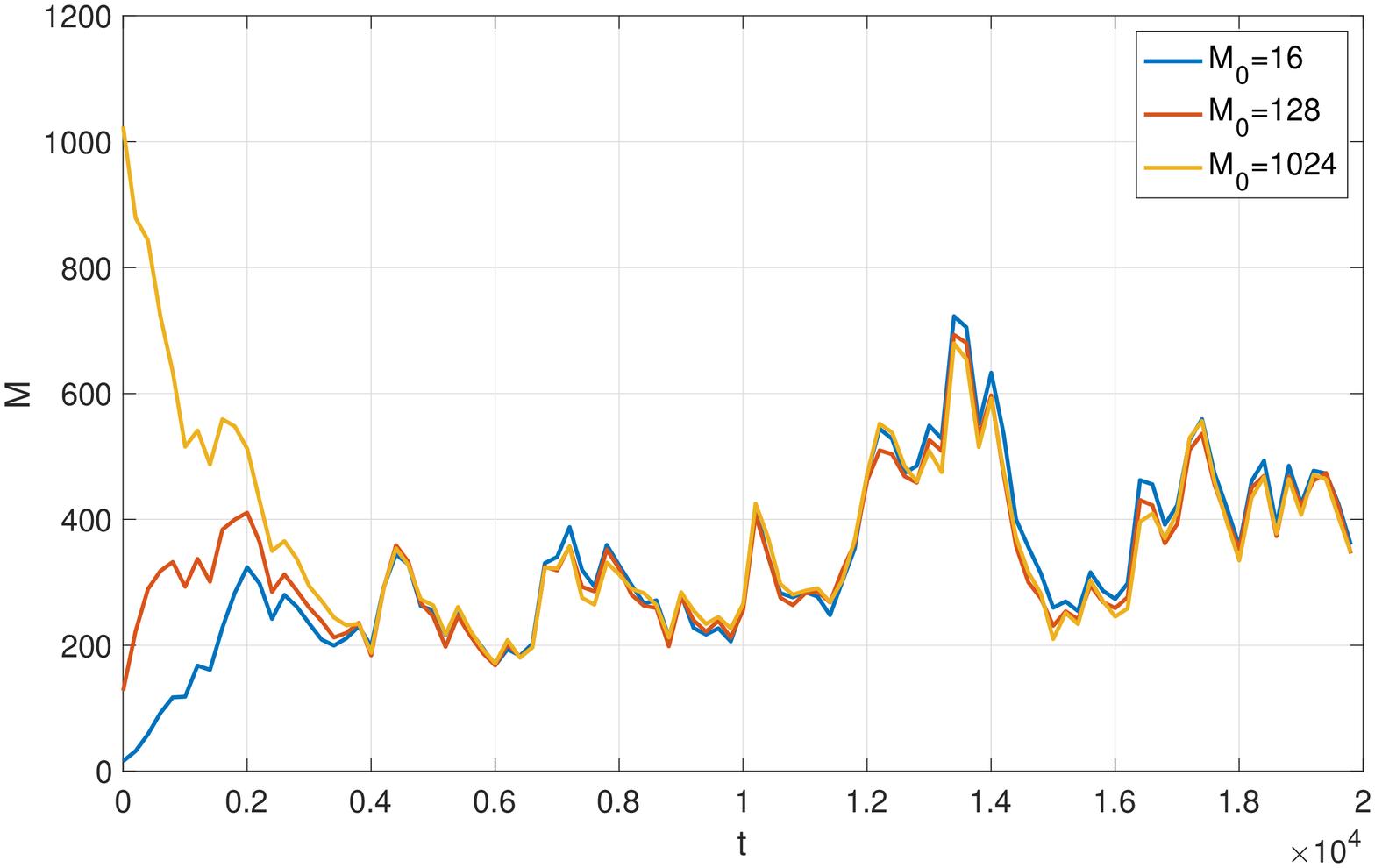}
} 
\centering
\subfigure[$K=9$]{
\includegraphics[width=0.49\textwidth]{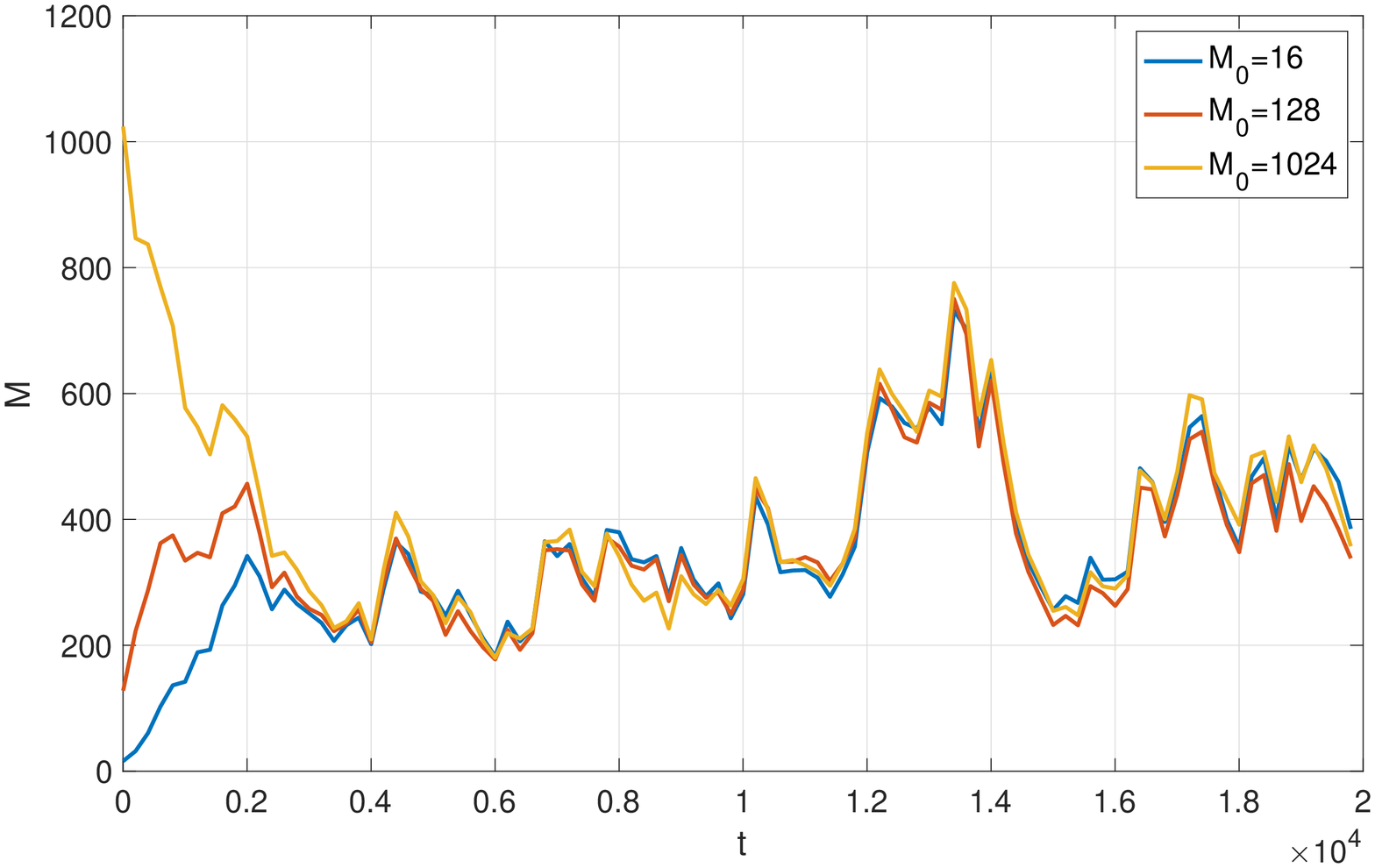}
} 
\caption{Averaged number of particles as a function of time for a fixed number of fictitious observations {$K = \{5,9\}$}.}
\label{fig_adaptive_M_vs_time_fixed_K}
\end{figure}

Figures \ref{fig_adaptive_M_vs_time_fixed_M0} and \ref{fig_adaptive_M_vs_time_fixed_K} show the averaged number of particles (over 100 independent runs) as a function of time. In Fig. \ref{fig_adaptive_M_vs_time_fixed_M0}, each subplot is obtained by fixing $M_0 \in \{16,128,1024\}$ and each line represents the evolution of the number of particles for each {$K\in\{ 2, 3, 5, 7, 9, 15, 20\}$}. We see that for most values $K$, the averaged number of particles is very similar. It is interesting to note that at some stages, the required number of particles is larger, and this is better detected with slightly larger values of $K$. In Fig. \ref{fig_adaptive_M_vs_time_fixed_K}, we show the same information, but now each subplot is obtained by fixing {$K\in\{        5  ,         9  \}$}, and each line represents the evolution of the number of particles for each initialization $M_0 \in \{16,128,1024\}$. We note that regardless the initial number of particles, after around $3,000$ time steps, the averaged number of particles is the same.

\begin{table*}[h]
\setlength{\tabcolsep}{2pt}
\def\marginwidth{1.5mm}
\begin{center}
\begin{tabular}{|l@{\hspace{\marginwidth}}|c@{\hspace{\marginwidth}}|c@{\hspace{\marginwidth}}|c@{\hspace{\marginwidth}}|c@{\hspace{\marginwidth}}|c@{\hspace{\marginwidth}}|c@{\hspace{\marginwidth}}|c@{\hspace{\marginwidth}}|c@{\hspace{\marginwidth}}|c@{\hspace{\marginwidth}}|c@{\hspace{\marginwidth}}|c@{\hspace{\marginwidth}}|c@{\hspace{\marginwidth}}|c@{\hspace{\marginwidth}}|c@{\hspace{\marginwidth}}|c@{\hspace{\marginwidth}}|c@{\hspace{\marginwidth}}|c@{\hspace{\marginwidth}}|c@{\hspace{\marginwidth}}|c@{\hspace{\marginwidth}}|c@{\hspace{\marginwidth}}|}

\hline
 {$M$ $/$ $K$} &2 &3 &5 &7 &10 &20 \\
\hline
\hline
   2 &$0.0006547$ & $5.571e-06$ & $1.891e-08$ & $2.466e-10$ & $7.979e-13$ & $0$   \\ 
   \hline
 4 &$0.0358$ & $0.009322$ & $0.001157$ & $0.0003728$ & $4.382e-05$ & $1.473e-07$   \\ 
 \hline
 16 &$0.3775$ & $0.3153$ & $0.2684$ & $0.2396$ & $0.2293$ & $0.1996$  \\ 
 \hline
 64 &$0.4681$ & $0.4354$ & $0.4288$ & $0.4285$ & $0.4256$ & $0.4117$  \\ 
 \hline
256 &$0.5511$ & $0.5495$ & $0.5667$ & $0.5801$ & $0.5759$ & $0.5781$  \\ 
\hline
1024 & $0.5793$ & $0.578$ & $0.5738$ & $0.5863$ & $0.5845$ & $0.6074$  \\ 
\hline
4096 &$0.5472$ & $0.5686$ & $0.5778$ & $0.5913$ & $0.5949$ & $0.593$   \\
\hline
\end{tabular}
\end{center}
\caption{ {\textbf{(Ex. of Section \ref{K_influence})} p-value of the uniformity test for different values of $K$ and $M$, averaged over $1000$ independent runs. For each run, the number of particles $M$ is fixed.} 
}
\label{table_pval_fixed_M}
\end{table*}

\begin{table*}[h]
\setlength{\tabcolsep}{2pt}
\def\marginwidth{1.5mm}
\begin{center}
\begin{tabular}{|l@{\hspace{\marginwidth}}|c@{\hspace{\marginwidth}}|c@{\hspace{\marginwidth}}||c@{\hspace{\marginwidth}}|c@{\hspace{\marginwidth}}|c@{\hspace{\marginwidth}}|c@{\hspace{\marginwidth}}|c@{\hspace{\marginwidth}}|c@{\hspace{\marginwidth}}|c@{\hspace{\marginwidth}}|c@{\hspace{\marginwidth}}|c@{\hspace{\marginwidth}}|c@{\hspace{\marginwidth}}|c@{\hspace{\marginwidth}}|c@{\hspace{\marginwidth}}|c@{\hspace{\marginwidth}}|c@{\hspace{\marginwidth}}|c@{\hspace{\marginwidth}}|c@{\hspace{\marginwidth}}|c@{\hspace{\marginwidth}}|c@{\hspace{\marginwidth}}|}
\hline
& & & \multicolumn{9}{|c| }{ $K$}\\
\cline{4-12}
 $T$ & $M_0$ & $W$ &2 &3 &5 &7 &9 &10 &11 &15 &20 \\
\hline  
\hline
$2\cdot 10^4$ & 16 & 50 & $153.47$ & $207.93$ & $231.86$ & $251.93$ & $247.18$ & $248.27$ & $251.18$ & $258.31$ & $272.79$ \\
$2\cdot 10^4$ & 128 & 50 & $151.82$ & $204.25$ & $231.2$ & $252.19$ & $246.66$ & $248.01$ & $255.4$ & $255.1$ & $279.91$ \\ 
$2\cdot 10^4$ & 1024 & 50 & $152.38$ & $209.92$ & $231.68$ & $248.69$ & $251.2$ & $246.17$ & $254.79$ & $251.57$ & $267.49$ \\  

\hline
$2\cdot 10^4$ & 16 & 200 & $303.02$ & $371.48$ & $417.23$ & $414.44$ & $434.48$ & $434.94$ & $433.03$ & $440.29$ & $438.54$ \\
$2\cdot 10^4$ & 128 & 200 & $304.23$ & $382.7$ & $399.5$ & $418.79$ & $421.21$ & $438.79$ & $436.32$ & $437.64$ & $439.61$ \\ 
$2\cdot 10^4$ & 1024 & 200 &$301.2$ & $366.35$ & $400.22$ & $432.99$ & $447.4$ & $434.83$ & $431.76$ & $444.38$ & $445.81$ \\ 
\hline
\end{tabular}
\end{center}
\caption{ \textbf{(Ex. of Section \ref{K_influence})} Averaged number of particles in the last {50} windows with $T=10^4$, averaged over {100} independent runs for different values of the window size $W\in\{50,200\}$ and initial number of particles $M_0 = \{16,128,1024\}$.}
\label{table_averaged_M}
\end{table*}

{\color{black}
\subsection{The three-dimensional Lorenz system}

Table \ref{table_lorenz_correlation} shows results of the Lorenz example described in  \cite[Section V-A]{elvira2017adapting} with fixed number of particles $M$. 
We show the MSE in the approximation of the posterior mean, averaged over $200$ runs. Again $r$ is the sample Pearson's correlation coefficient, using the whole sequence of statistics $\{ a_{K,M,t}\}_{t=1}^T$ with a lag $\tau = 1$, and p-val is the p-value of the Pearson's $\chi^2$ test for assessing the uniformity of the same set. Similar conclusions as in the previous example can be obtained.

\begin{table*}
\setlength{\tabcolsep}{2pt}
\def\marginwidth{1.5mm}
\begin{center}
{\color{black}
\begin{tabular}{|l@{\hspace{\marginwidth}}|c@{\hspace{\marginwidth}}|c@{\hspace{\marginwidth}}|c@{\hspace{\marginwidth}}|c@{\hspace{\marginwidth}}|c@{\hspace{\marginwidth}}|c@{\hspace{\marginwidth}}|c@{\hspace{\marginwidth}}|c@{\hspace{\marginwidth}}|c@{\hspace{\marginwidth}}|c@{\hspace{\marginwidth}}|c@{\hspace{\marginwidth}}|c@{\hspace{\marginwidth}}|c@{\hspace{\marginwidth}}|c@{\hspace{\marginwidth}}|c@{\hspace{\marginwidth}}|c@{\hspace{\marginwidth}}|c@{\hspace{\marginwidth}}|c@{\hspace{\marginwidth}}|c@{\hspace{\marginwidth}}|c@{\hspace{\marginwidth}}|}

\hline
\cline{2-4}
Fixed $M$  &8 &16 &32 &64 &128 &256 & 512 & 1024  & 2048 & 4096 & 8192 & 16384 \\
\hline
\hline
 MSE  &105.63 &  75.56   &40.19  & 15.69   & 5.90 &    2.90 &   1.77  &  1.55  &1.53    &1.52   & 1.52  &  1.52 \\
\hline
$\hat R(1)$ & 0.6927  &  0.4939  &  0.2595  &  0.1132  &  0.0463  &  0.0273   & 0.0210  &  0.0190  & 0.0195  &  0.0151  &  0.0151   & 0.0192 \\
\hline
 p-val  &0.0393  &  0.1276   & 0.2923    &0.4279  &  0.4823  &  0.5016  &  0.5117   & 0.5106 &   0.4998 &   0.5141   & 0.5040 & 0.5181\\
\hline
\end{tabular}
}
\end{center}
\caption{{Lorenz Model:  $\Delta=10^{-3}$, $T_{obs}=200\Delta$, $\sigma^2=0.5$. Algorithm details: $W=20$, $K=7$. MSE in the approximation of the posterior mean, the averaged 
$\hat R(1)$, and the averaged p-value of the Pearson's chi-square test on the uniformity on $\mathcal{S}_t$.}
\label{table_lorenz_correlation}}
\end{table*}

}

\subsection{Behavior of $A_{K,M,t}$ and its relation with $B_{M,t}$}
\label{sec_numerical_1}
\begin{figure*}[h] 
\centering
\includegraphics[width=0.75\textwidth]{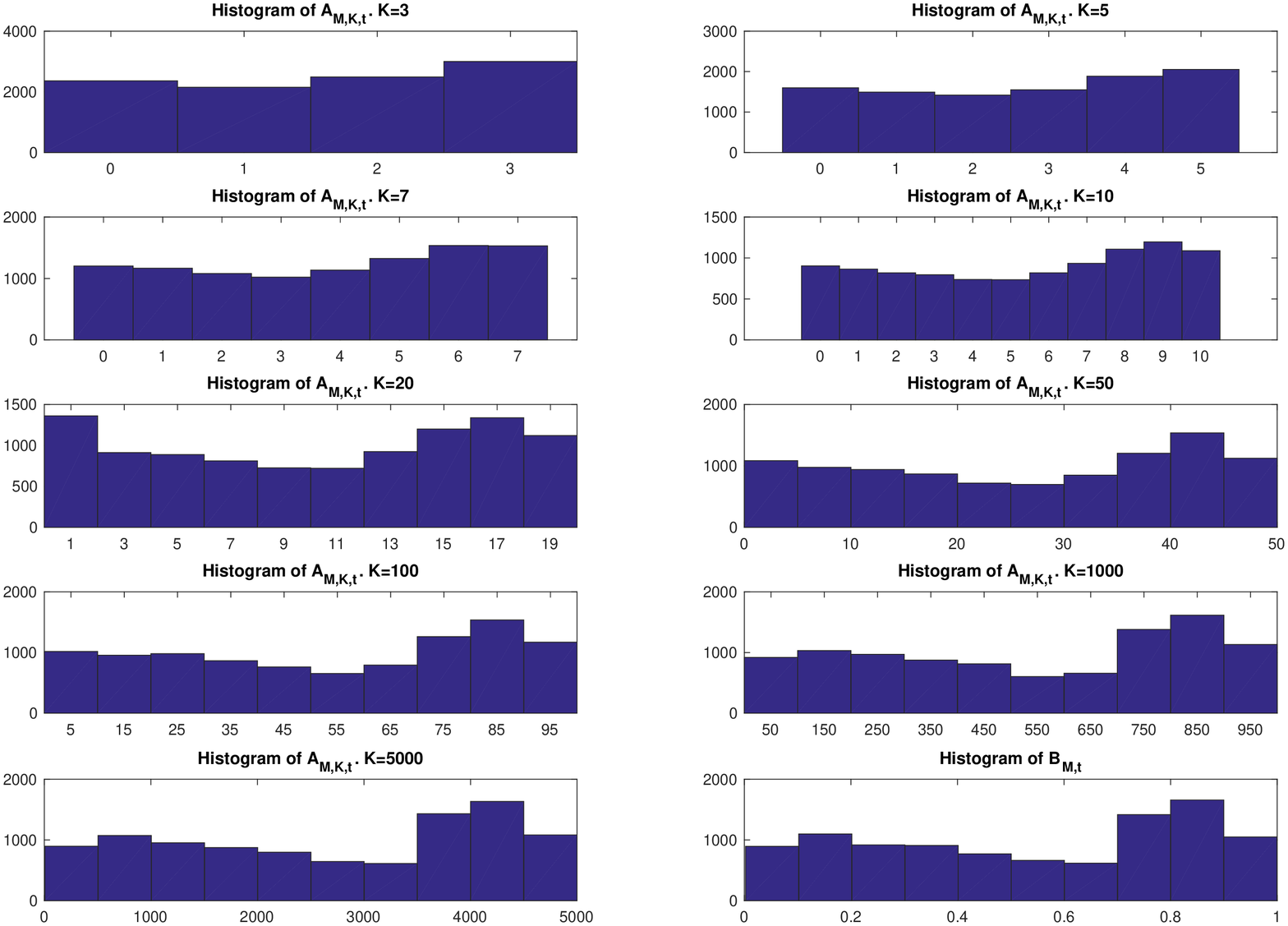}
\caption{\textbf{(Ex. of Section \ref{sec_numerical_2})} Histograms of the r.v.  $A_{K,M,t}$ and $B_{M,t}$ in the stochastic growth model with fixed $M=2^{14}$. The histograms accumulate $T=100$ time steps in $100$ independent runs. }
\label{hist_A_B_sgm}
\end{figure*}

In Fig. \ref{hist_A_B_sgm}, {we show the histograms of $A_{K,M,t}$ and $B_{M,t}$ for} the stochastic growth model described in \eqref{sgm_state}-\eqref{sgm_obs}. We set $K\in\{3,5,7,10,20,50,100,1000,5000\}$. The 
BF is with fixed $M=2^{14}$. 
When $K$ grows, the pmf 
seems to converge to the pdf of $B_{M,t}$. 

In Table, \ref{table_dist_A_B_sgm} we show the averaged absolute error (distance) between the realizations of r.v.'s  $A_{K,M,t}/K$ and $B_{M,t}$ for the stochastic growth model with fixed $M=2^{14}$. The results are averaged over $T=100$ time steps in $100$ independent runs. It is clear that when $K$ grows, the deviation between both r.v.'s, {which take values in $(0,1)$}, decreases. 
Thus, for $K=5000$, the difference is on average $0.43\%$.
\begin{table*}[h]
\setlength{\tabcolsep}{2pt}
\def\marginwidth{1.5mm}
\begin{center}
\begin{tabular}{|l@{\hspace{\marginwidth}}|c@{\hspace{\marginwidth}}|c@{\hspace{\marginwidth}}|c@{\hspace{\marginwidth}}|c@{\hspace{\marginwidth}}|c@{\hspace{\marginwidth}}|c@{\hspace{\marginwidth}}|c@{\hspace{\marginwidth}}|c@{\hspace{\marginwidth}}|c@{\hspace{\marginwidth}}|c@{\hspace{\marginwidth}}|c@{\hspace{\marginwidth}}|c@{\hspace{\marginwidth}}|c@{\hspace{\marginwidth}}|c@{\hspace{\marginwidth}}|c@{\hspace{\marginwidth}}|c@{\hspace{\marginwidth}}|c@{\hspace{\marginwidth}}|c@{\hspace{\marginwidth}}|c@{\hspace{\marginwidth}}|c@{\hspace{\marginwidth}}|}

\hline
 $K$ &2 &3 &5 &7 &10 &20 &50 &100 &1000 &5000\\
\hline
\hline
$\big|B_{M,t} - \frac{A_{M,K,t}}{K}\big|$ &0.2254 & 0.1836  & 0.1409   & 0.1183   & 0.0987   & 0.0696  &  0.0435   & 0.0305  &  0.0097  &  0.0043 \\
\hline
\end{tabular}
\end{center}
\caption{\textbf{(Ex. of Section \ref{sec_numerical_2})} Averaged absolute error (distance) between the realizations of the r.v.'s  $A_{K,M,t}$ and $B_{M,t}$ for the stochastic growth model $M=2^{14}$. The results are averaged over $T=100$ time steps in $100$ independent runs.}
\label{table_dist_A_B_sgm}
\end{table*}

\vspace{-0.5cm}
\subsection{Forgetting property in the block-adaptive bootstrap particle filter}
\label{sec_numerical_2}
In this section, we assess the approximation errors when the block-adaptive BF increases the number of particles. To that end, we run two specific state-space models where, in the first half of time steps, the number of particles is set to $M_1$ while, in the second half, the number of particles is $M_2>M_1$. We then compute the MSE {of predicted observations}  (in the last quarter of the time steps), {and we compare it with the standard BF with $M_2$ particles used from the beginning.} 

Table \ref{table_lgm_hh} shows the MSE of a BF run on the linear Gaussian model described by 
\beqa
x_t&=&a x_{t-1} +  u_t,\\
y_t&=& x_t + v_t,
\eeqa
with $T=1000$, {$\sigma_u = \sqrt{0.5}$, $\sigma_v = 1$,} and $a=0.9$. We simulate {one} example with $M_1 = 100$ and $M_2 = 1000$ (left side of the table), and {another} one with $M_1 = 1000$ and $M_2 = 10000$ (right side of the table). In both cases, we are able to show that the BF achieves in the last quarter of time steps (from $t=750$ to $t=T=1000$) the same MSE as if the largest number of particles was set {at the very beginning}.

Table \ref{table_sgm_hh} {presents analogous results} 
for the stochastic growth model {described in the first experiment}, with  $T=1000$, $\sigma_x = 1$, $\sigma_y^2 = 0.1$, and $\phi=0.4$. Now we simulate the BF with the following pairs of number of particles $(M_1,M_2) \in \{(50,1000),(200,4000),(1000,20000)\}$. {We arrive at the same conclusions.} 

\begin{table*}[h]
\setlength{\tabcolsep}{2pt}
\def\marginwidth{1.5mm}
\begin{center}
\begin{tabular}{|l@{\hspace{\marginwidth}}||c|c|c||c|c|c|} 
\hline
\cline{2-4}
 $M_1$  &\multirow{2}{*}{100} &\multirow{2}{*}{1000}  &100 &  \multirow{2}{*}{1000} & \multirow{2}{*}{10000}&  1000 \\ 
\cline{1-1}\cline{4-4}\cline{7-7}
$M_2$  & &  &1000 &  &  &  10000 \\ 
\hline
\hline
 MSE (last $T/4$) & $8.90\dot 10^{-3}$ &  $9.02\dot 10^{-4}$  &   $8.99\dot 10^{-4}$  &    $9.02\dot 10^{-4}$  & $8.93 \dot 10^{-5}$ & $8.69 \dot 10^{-5}$\\ 
\hline
\end{tabular}
\end{center}
\caption{Linear Gaussian Model: $T=1000$, {$\sigma_x = \sqrt{0.5}$, $\sigma_y = 1$,} $a=0.9$. $M_1$ particles for $t \in \{1,...,\frac{T}{2}\}$ and $M_2$ particles for $t \in \{\frac{T}{2}+1,...,T\}$.}
\label{table_lgm_hh}
\end{table*}

\begin{table*}[h]
\setlength{\tabcolsep}{2pt}
\def\marginwidth{1.5mm}
\begin{center}
\begin{tabular}{|l@{\hspace{\marginwidth}}||c|c|c||c|c|c||c|c|c|}
\hline
\cline{2-4}
 $M_1$  &\multirow{2}{*}{50} &\multirow{2}{*}{1000}  &50 &  \multirow{2}{*}{200} & \multirow{2}{*}{4000}& 200 &  \multirow{2}{*}{1000} & \multirow{2}{*}{20000}& 1000  \\ 
\cline{1-1}\cline{4-4}\cline{7-7}\cline{10-10}
$M_2$  & &  &1000 &  &  &  4000 & & & 20000 \\ 
\hline
\hline
 MSE (last $T/4$) & $16.69$ &  $1.493$  & $1.564$  &    $4.815$  & $1.386$ & $1.374$ & $1.494$ & $1.348$& $1.335$\\ 
\hline
\end{tabular}
\end{center}
\caption{Stochastic Growth Model: $T=1000$, $\sigma_x = 1$, $\sigma_y = 0.1$, $\phi=0.4$. $M_1$ particles for $t \in \{1,...,\frac{T}{2}\}$ and $M_2$ particles for $t \in \{\frac{T}{2}+1,...,T\}$.}
\label{table_sgm_hh}
\end{table*}

\section{Summary and conclusions}

\cbrown{ 
In this paper, we have provided new methodological, theoretical and numerical results on the performance of particle filtering algorithms with an adaptive number of particles. We have looked into a class of PFs that update the number of particles periodically, at the end of observations blocks of a prescribed length. Decisions on whether to increase or decrease the computational effort are automatically made based on predictive statistics which are computed by generating \textit{fictitious observations}, i.e., particles in the observation space. For this type of algorithms, we have proved that:
\begin{itemize}
    \item[(a)] The error bounds for the adaptive PF depend on the current number of particles (say, $M_n$) and the dependence on former values (say $M_{n-1}, M_{n-2}, \dots$) decays exponentially with the block length. This results holds under standard assumptions on the Markov kernel of the state-space model (as discussed in \citet{DelMoral04,Kunsch05} and others). This result, which does not follow from classical convergence theorems for Monte Carlo filters, implies that one can effectively tune the performance of the PF by adapting the computational effort.  
    \item[(b)] Convergence of the predictive statistics used for making decisions on the adaptation of the computational effort implies convergence of the PF itself. To be specific, we have proved that if the predictive statistics computed with $K$ fictitious observations attain a uniform distribution then the true filtering distribution and its particle approximation have $K$ common moments. This result can be understood as a converse to the convergence theorem introduced in \citet{elvira2017adapting}. It guarantees that assessing the convergence of the PF using the proposed predictive statistics is a sound approach (the ``more uniform'' the predictive statistics, the better the PF general performance). 
\end{itemize}
In addition to the theoretical analysis, we have carried out an extensive computer simulation study. On one hand, the numerical results have corroborated the theoretical results, e.g., by showing how increasing the number of particles directly improves the performance (and past errors are forgotten), or how increasing the number of fictitious observations $K$ (or the block size) leads to a higher computational effort and more accurate estimators. We have also shown that the proposed block-adaptive algorithms are stable w.r.t. the initial number of particles. 
}

\cbrown{ 
Overall, the proposed class of algorithms is easy to implement and can be used with different versions of the PF. It also enables the automatic, online tuning of the computational complexity without time-consuming (and often unreliable) trial-and-error procedures. 
}

\appendix

\section{Proof of Theorem \ref{thConvergence}} \label{apConvergence}

The argument of the proof relies on the following lemma:
\begin{Lema} \label{lmContractPU}
If Assumptions \ref{asMixing} and \ref{asLikelihood} hold, the composition of $r \ge 1$ consecutive PU operators satisfies the inequality
\beqa
\sup_{|f|\le 1} \left|
	\left(f, \Psi_{t|t-r}(\alpha)\right) - \left(f, \Psi_{t|t-r}(\beta)\right)
\right| &\le& \nn\\
\frac{
	2\left(
		1 - \varepsilon_m^2\gamma^{m-1}
	\right)^{ \lfloor \frac{ r }{ m } \rfloor }
}{
	\gamma^m \varepsilon_m
} \sup_{|\tilde f| \le 1} \left|
	(\tilde f,\alpha)-(\tilde f,\beta)
\right|
\nn
\eeqa
for any probability measures $\alpha,\beta \in \mP(\mX)$.
\end{Lema}
The proof of Lemma \ref{lmContractPU} follows immediately from Propositions 4.3.3 and 4.3.7 in \cite{DelMoral04}. This is now a classical result that has been exploited for many particle-based algorithms (see, e.g., \cite{Gupta15} or \cite{crisan2017uniform}).

Recall that $t_n = \sum_{j^0}^n W_j$ is the last time instant of the $n$th block of observations. For any bounded real test function $f:\mX\mapsto\Real$, the approximation error $(f,\pi_{t_n}^{M_n})-(f,\pi_{t_n})$ can be written in terms of one-step-ahead differences using the ``telescope'' decomposition 
\beq
(f,\pi_{t_n}^{M_n})-(f,\pi_{t_n}) = \sum_{k=0}^{W_n-2} D_k(M_n) + D(M_n,M_{n-1}) + E(M_{n-1}),
\label{eqTelescope}
\eeq
where 
\beqa
D_k(M_n) &:=& \left(
	f,\Psi_{t_n|t_n-k}\left(\pi_{t_n-k}^{M_n}\right)
\right) \nn \\
&& - \left(
	f,\Psi_{t_n|t_n-k}\left(
		\Psi_{t_n-k}\left(\pi_{t_n-k-1}^{M_n}\right)
	\right)
\right) \nn
\eeqa
are the local (one step) differences in the $n$-th window,
\beqa
D(M_n,M_{n-1}) &=& \left(
	f, \Psi_{t_n|t_n-W_n+1}\left( \pi_{t_n-W_n+1}^{M_n} \right) 
\right) \nn\\
&& - \left(
	f, \Psi_{t_n|t_n-W_n+1}\left(
		\Psi_{t_n-W_n+1}\left( \pi_{t_n-W_n}^{M_{n-1}} \right)
	\right) 
\right) \nn
\eeqa
is the one-step difference between the last time instant of window $n-1$ and the first time instant of window $n$, and
\beqa
E(M_{n-1}) &=& \left(
	f, \Psi_{t_n|t_n-W_n}\left( \pi_{t_n-W_n}^{M_{n-1}} \right) 
\right) \nn\\
&& - \left(
	f, \Psi_{t_n|t_n-W_n}\left( \pi_{t_n-W_n} \right) 
\right) \nn
\eeqa
is the approximation error inherited from window $n-1$. Using Lemma \ref{lmContractPU} and writing $t_{n-1}=t_n-W_n$ for conciseness, we readily find that for any test function $f$, with $\|f\|_\infty\le 1$, the different terms on the rhs of \eqref{eqTelescope} can be upper bounded as
\beqa
|D_k(M_n)| &\le& \frac{
	2\left(
		1 - \varepsilon_m^2\gamma^{m-1}
	\right)^{ \lfloor \frac{ k }{ m } \rfloor }
}{
	\gamma^m \varepsilon_m
} \nn \\
&& \times \sup_{|\tilde f| \le 1} \left|
	(\tilde f,\pi_{t_n-k}^{M_n})-\left(\tilde f,\Psi_{t_n-k}\left( \pi_{t_n-k-1}^{M_n}\right)\right)
\right|, \nn \\ \label{eqDif_k}
\eeqa
\beqa
| D(M_n,M_{n-1}) | \le \frac{
	2\left(
		1 - \varepsilon_m^2\gamma^{m-1}
	\right)^{ \lfloor \frac{ W_n-1 }{ m } \rfloor }
}{
	\gamma^m \varepsilon_m
} && \nn \\
\times \sup_{|\tilde f| \le 1} \left|
	(\tilde f,\pi_{t_{n-1}+1}^{M_n})-\left(\tilde f,\Psi_{t_{n-1}+1}\left( \pi_{t_{n-1}}^{M_{n-1}}\right)\right)
\right|,&& \label{eqDif_W}
\eeqa
and
\beqa
|E(M_{n-1})| &\le& \frac{
	2\left(
		1 - \varepsilon_m^2\gamma^{m-1}
	\right)^{ \lfloor \frac{ W_n }{ m } \rfloor }
}{
	\gamma^m \varepsilon_m
} \nn \\
&& \times \sup_{|\tilde f| \le 1} \left|
	(\tilde f,\pi_{t_{n-1}}^{M_{n-1}})-\left(\tilde f,\pi_{t_{n-1}}^{M_{n-1}} \right)
\right|.
\label{eqEn-1}
\eeqa
Moreover, the measures $\pi_{t_n-k}^{M_n}$ are importance sampling approximations of $\Psi_{t_n-k}\left( \pi_{t_n-k-1}^{M_n} \right)$, for $k=0, ..., W_n-2$. As a consequence, it can easily be shown that that there is a constant $0<c<\infty$, independent of $M_n$, $t_n$ and $k$, such that
\beq
\left\|
	(f,\pi_{t_n-k}^{M_n}) - \left( f,\Psi_{t_n-k}\left( \pi_{t_n-k-1}^{M_n}\right) \right)
\right\|_p \le \frac{\|f\|_\infty c}{\sqrt{M_n}}
\label{eqMC1}
\eeq
for any bounded test function $f$ and any $p>1$. The same result also holds for $\pi_{t_{n-1}+1}^{M_n}$ and $\Psi_{t_{n-1}+1}\left( \pi_{t_{n-1}}^{M_{n-1}} \right)$, namely,
\beq
\left\|
	(f,\pi_{t_{n-1}+1}^{M_n}) - \left( f,\Psi_{t_{n-1}+1}\left( \pi_{t_{n-1}}^{M_{n-1}}\right) \right)
\right\|_p \le \frac{\|f\|_\infty c}{\sqrt{M_n}}.
\label{eqMC2}
\eeq 

Finally, if we apply Minkowsky's inequality in Eq. \eqref{eqTelescope} and then combine the bounds \eqref{eqDif_k} and \eqref{eqDif_W} with \eqref{eqMC1} and \eqref{eqMC2}, respectively, we obtain, for any $p\ge 1$
\beqa
&\sup_{|f|\le 1} \left| (f,\pi_{t_n}^{M_n}) - (f,\pi_{t_n}) \right|_p \le
\sum_{k=0}^{W_n-1} \frac{
	2c\left(
		1 - \varepsilon_m^2\gamma^{m-1}
	\right)^{ \lfloor \frac{ k }{ m } \rfloor }
}{
	\gamma^m \varepsilon_m \sqrt{M_n}
}& \nn\\
& + \frac{
	2\left(
		1 - \varepsilon_m^2\gamma^{m-1}
	\right)^{ \lfloor \frac{ W_n }{ m } \rfloor }
}{
	\gamma^m \varepsilon_m
} \sup_{|f| \le 1} \left\|
	\left( f,\pi_{t_{n-1}}^{M_{n-1}} \right) - \left( f,\pi_{t_{n-1}} \right)
\right\|_p.&\nn\\
\label{eqAlmostThere}
\eeqa
The proof is complete by pointing out that, taking $W_n\rw\infty$ in the sum $\sum_{k=0}^{W_n-1}(\cdot)$ on the rhs of \eqref{eqAlmostThere}, we arrive at
\beq
\sum_{k=0}^{W_n-1} \frac{
	2c\left(
		1 - \varepsilon_m^2\gamma^{m-1}
	\right)^{ \lfloor \frac{ k }{ m } \rfloor }
}{
	\gamma^m \varepsilon_m \sqrt{M_n}
} \le \frac{
	mC
}{
	\varepsilon_m^3 \gamma^{2m-1} \sqrt{M_n}
},
\eeq
where $C=2c$. 

\section{Proof of Theorem \ref{thMoments}} \label{appendix_lemma2}

First, we express the pmf $\hat \mbbQ_{K,t}$ of the r.v. $\hat A_{K,t}$ as a function of the predictive pdf of the observations. Specifically, we note that $\hat \mbbQ_{K,t}(n)$ is the probability that exactly $n$ fictitious observations are smaller than the actual $y_t$. Hence, $\forall n\in\{0,...,K\}$, 
\begin{align}
&\hat \mbbQ_{K,t}(n)  = \int_{-\infty}^\infty \mathbb{P}(\hat A_{K,t}=n|y_t=z)p_t(z)dz \nonumber \\
&= \int_{-\infty}^\infty  {{K}\choose{n}} \hat F_t(z)^n \left( 1-\hat F_t(z)\right)^{K-n}p_t(z)dz \nonumber\\
&=    {{K}\choose{n}} \int_{-\infty}^\infty \hat F_t(z)^{n} \left( \sum_{i=0}^{K-n}{{K-n}\choose{i}} (-1)^{{i}} \hat F_t(z)^{i}\right)p_t(z)dz \nonumber \\
&=    {{K}\choose{n}}  \sum_{i=0}^{K-n}\int_{-\infty}^\infty {{K-n} \choose{i}} (-1)^{{i}} \hat F_t(z)^{n+i} p_t(z)dz,
\label{pmf_different_n}
\end{align}
where $\hat F_t(z)$ is the approximation of the cdf of the predictive observation evaluated at $z$. Note that, as a consequence, $\hat F_t(z)$ is also the probability of a single fictitious observation being smaller than $z$.

Recall that we want to prove that \eqref{eq_lemma_2} when $\hat A_{K,t}$ is uniform (and, as a consequence, $\hat \mbbQ_{K,t}(n) = \frac{1}{K+1}$ for every $n \in \{0, \ldots, K \}$). We apply an induction argument in $n$. The case for $n=K$ is obvious by rewriting \eqref{pmf_different_n} as
\begin{align}
\hat \mbbQ_{K,t}(K) &= \int_{-\infty}^\infty \hat F_t(z)^{K} p_t(z)dz = ( \hat F_t^{K},p_t),
\nn
\end{align}
hence $( \hat F_t^{K},p_t) = \frac{1}{K+1}$.
Next, we assume that for a specific $n \in \{1,...,K \}$, the identity \eqref{eq_lemma_2} holds for all $i \in \{n,...,K \}$ and then aim at proving that it also holds for $n-1$. Let us write the pmf of \eqref{pmf_different_n}  at $n-1$ as
{\small
\begin{align}
&\hat \mbbQ_{K,t}(n-1)=  \nonumber\\
&=  {{K}\choose{n-1}}  \sum_{i=0}^{K-n+1}\int_{-\infty}^\infty {{K-n+1} \choose{i}} (-1)^{{i}} \hat F_t(z)^{n+i-1} p_t(z)dz \nonumber\\
&=  {{K}\choose{n-1}}  \left[
	\sum_{i=1}^{K-n+1} {{K-n+1} \choose{i}} \frac{(-1)^{{i}}}{n+i} + \int_{-\infty}^\infty  \hat F_t(z)^{n-1} p_t(z)dz
\right] \nonumber \\
&=  {{K}\choose{n-1}}  \frac{1}{n} \left( { \sum_{i=0}^{K-n+1} {{K-n+1} \choose{i}} (-1)^{{i}} \frac{n}{n+i}} -1 \right) \nonumber \\
&\quad + {{K}\choose{n-1}} \int_{-\infty}^\infty  \hat F_t(z)^{n-1} p_t(z)dz \nonumber \\
&=  {{K}\choose{n-1}}  \left[
	\frac{1}{n} \left( {\frac{1}{ {{K+1}\choose{K-n+1}}}} -1 \right) + \int_{-\infty}^\infty  \hat F_t(z)^{n-1} p_t(z)dz
\right],
\label{pmf_different_n_1}
\end{align}
where the second identity is obtained replacing all the integrals (except the one corresponding to $n-1$) using the induction hypothesis; for the third equality, we split the series between the terms $i>0$ and the term $i=1$, and in the fourth equation, we substitute the series using identity (1.41) in \cite{gould1972combinatorial}. Once again, since $\hat A_{K,t}$ is uniform, we have $\hat \mbbQ_{K,t}(n-1) = \frac{1}{K+1}$, hence we rewrite \eqref{pmf_different_n_1} as 
\beqa
\frac{1}{K+1} &=& {{K}\choose{n-1}}  \frac{1}{n} \left( \frac{1}{ {{K+1}\choose{K-n+1}}} -1 \right) \nn\\
&& + {{K}\choose{n-1}} \int_{-\infty}^\infty  \hat F_t(z)^{n-1} p_t(z)dz, \nn
\eeqa
where, precisely, $\int_{-\infty}^\infty  \hat F_t(z)^{n-1} p_t(z)dz = (\hat F_t^{n-1},p_t)$. If we simply solve for the latter integral and simplify, we arrive at 
{\small
\begin{align}
 &(\hat F_t^{n-1},p_t) =  \frac{1}{{{K}\choose{n-1}}}  \Bigg(   \frac{1}{K+1} - {{K}\choose{n-1}}  \frac{1}{n} \left( \frac{1}{ {{K+1}\choose{K-n+1}}} -1 \right) \Bigg) \nonumber \\
 &= \frac{(n-1)!(K-n+1)!}{K!}\frac{1}{K+1} 
 -\frac{1}{n} \left(   \frac{(K-n+1)!n!}{(K+1)!} - 1 \right) \nonumber \\
  &= \frac{(n-1)!(K-n+1)!}{(K+1)!} 
  - \frac{(K-n+1)!(n-1)!}{(K+1)!}  + \frac{1}{n} \nonumber \\
  & = \frac{1}{n}. \qed
\end{align}
}

\bibliographystyle{spbasic}



\end{document}